\documentclass[aip,jcp,%
 twocolumn,
 amsmath,
 amssymb,
 reprint,
 citeautoscript,
 float
]{revtex4-1}

\usepackage{graphicx}%
\usepackage{array,multirow}
\usepackage{amsmath} %
\usepackage{amssymb} %
\usepackage{braket}
\usepackage{mathtools} %
\usepackage{float}

\DeclarePairedDelimiter\sqbrak{ [ }{ ] }
\DeclarePairedDelimiter\paran{ ( }{ ) }

\newcommand*{\gA}{\boldsymbol{\gamma}^{\text{A}}} %
\newcommand*{\gB}{\boldsymbol{\gamma}^{\text{B}}} %
\newcommand*{\newgA}{\tilde{\boldsymbol{\gamma}}^{\text{A}}} %
\newcommand*{\vemb}{\mathbf{v}_\text{emb}} %
\newcommand*{\vembAB}{\mathbf{v}_\text{emb} \left[ \gA,\gB \right]} %
\newcommand*{\vembABtrun}{\mathbf{v}^{\text{trun}}_\text{emb} \left[ \gA,\gB \right]} %
\newcommand*{\dA}{\tilde{\mathbf{d}}^{\text{A}}}
\newcommand*{\dAtrun}{\tilde{\mathbf{d}}^{\text{A,trun}}}

\newcommand*{\citen}[1]{%
  \begingroup
    \romannumeral-`\x %
    \setcitestyle{numbers}%
    \cite{#1}%
  \endgroup   
}

\begin{document}

\title{Analytical Gradients for Projection-Based Wavefunction-in-DFT Embedding} %

\author{Sebastian J. R. Lee}
\affiliation{Division of Chemistry and Chemical Engineering, California Institute of Technology, Pasadena, California 91125, United States}

\author{Feizhi Ding}
\affiliation{Division of Chemistry and Chemical Engineering, California Institute of Technology, Pasadena, California 91125, United States}

\author{Frederick R. Manby}
\affiliation{Centre for Computational Chemistry, School of Chemistry, University of Bristol, Bristol BS8 1TS, United Kingdom}

\author{Thomas F. Miller III}
\email[]{tfm@caltech.edu}
\affiliation{Division of Chemistry and Chemical Engineering, California Institute of Technology, Pasadena, California 91125, United States}

\date{\today}

\begin{abstract}
Projection-based embedding provides a simple, robust, and accurate approach for describing a small part of a chemical system at the level of a correlated wavefunction method while the remainder of the system is described at the level of density functional theory. 
Here, we present the derivation, implementation, and numerical demonstration of analytical nuclear gradients for projection-based wavefunction-in-density functional theory (WF-in-DFT) embedding.
The gradients are formulated in the Lagrangian framework to enforce orthogonality, localization, and Brillouin constraints on the molecular orbitals.
An important aspect of the gradient theory is that WF contributions to the total WF-in-DFT gradient can be simply evaluated using existing WF gradient implementations without modification.
Another simplifying aspect is that Kohn-Sham (KS) DFT contributions to the projection-based embedding gradient do not require knowledge of the WF calculation beyond the relaxed WF density.
Projection-based WF-in-DFT embedding gradients are thus easily generalized to any combination of WF and KS-DFT methods. 
We provide numerical demonstration of the method for several applications, including calculation of a minimum energy pathway for a hydride transfer in a cobalt-based molecular catalyst using the nudged-elastic-band method at the CCSD-in-DFT level of theory, which reveals large differences from the transition state geometry predicted using DFT. 
\end{abstract}

\maketitle %

\section{Introduction}
The theoretical description of many chemical processes  demands  accurate, \textit{ab initio} electronic structure theories.
However, the study of complex reactive processes, including those arising in inorganic and enzyme catalysis, gives rise to the need for a compromise between accuracy and the ability to complete the computation in a reasonable amount of time.
For systems in which the complicated chemical rearrangements (e.g. bond breaking and forming) occurs in a local spatial region, 
an effective strategy for balancing accuracy and computational cost is to employ one of various multiscale embedding strategies \cite{Kitaura1999,Deev2005,Collins2006,Fedorov2007,Elliott2009,Goodpaster2010b,Huang2011,Goodpaster2011a,Manby2012,Goodpaster2012,Gordon2012,Knizia2012,Barnes2013,Goodpaster2014,Neuhauser2014,Barnes2015,Fornace2015,Bennie2015,Stella2015,Huo2016,Bennie2016,Pennifold2017,Zhang2018c,Chapovetsky2018,Muhlbach2018}. %
Generally, embedding methodologies hinge on the condition that a system can be efficiently partitioned into a local subsystem that demands a high-level treatment and an environment that can be treated with a lower (and computationally less expensive) level of theory.

The current paper focuses on projection-based embedding, \cite{Manby2012,Lee2019} a DFT-based embedding theory in which subsystem partitioning is performed in terms of localized Kohn-Sham (KS) molecular orbitals (LMOs).
The method describes subsystem interactions at the level of KS and allows for the partitioning of the subsystems across covalent and even conjugated bonds, and it enables the use of relatively small subsystem sizes for an embedded WF description. 
A recent review of projection-based WF-in-DFT embedding is available in Ref.~\citenum{Lee2019}. 

Projection-based embedding has proven to be a useful tool in a wide range of chemical contexts including transition-metal complexes \cite{Stella2015,Huo2016,Chapovetsky2018,Welborn2018}, protein active sites \cite{Bennie2016,Zhang2018c}, excited states \cite{DeLimaBatista2017,Bennie2017,Chen2019} and condensed phase systems \cite{Barnes2015}, among others \cite{Parrish2015,Libisch2017,Yao2017,Meitei2017,Chulhai2018,Lin2018,Bockers2018}. 
The development of analytical nuclear gradients for projection-based embedding will expand its applicability to include geometry optimization, transition state searches, and potentially \textit{ab initio} molecular dynamics.
Analytical nuclear gradients already exist for a number of other embedding methodologies, including the incremental molecular fragmentation method \cite{Hesselmann2018}, fragment molecular orbital method \cite{Nagata2012,Nakata2013,Brorsen2014}, quantum mechanics/molecular mechanics (QM/MM) \cite{Warshel1976,Field1990,Lin2007}, ONIOM \cite{Dapprich1999a,Hratchian2008,Mayhall2010,Hratchian2011}, embedded mean-field theory (EMFT) \cite{Fornace2015,entos}, frozen density embedding \cite{Duak2007,Heuser2016,Heuser2017,Heuser2018}, and subsystem DFT \cite{Kovyrshin2016a,Schluns2017,Klahr2018}. 
However, the projection-based approach provides a number of advantages for WF-in-DFT embedding calculations and leads to a distinct analytical gradient theory, which we derive and numerically demonstrate in several applications.

In section \ref{sec:energy} we outline projection-based WF-in-DFT embedding and in section \ref{sec:WFinDFT_gradient} we provide the derivation of its analytical nuclear gradients.
Section \ref{sec:results} numerically validates the analytical nuclear gradient theory and its implementation in Molpro\cite{MOLPRO} via comparison with finite difference calculations, as well as presenting results for optimizing geometries in benchmark systems and the calculation of a minimum energy profile for an organometallic reaction using the nudged-elastic-band (NEB) method.
We additionally provide the analytical nuclear gradient theory for WF-in-DFT embedding with atomic orbital (AO) truncation \cite{Bennie2015} in Appendices \ref{appendix:AO_energy} and \ref{appendix:WFinDFT_gradient_AO}.

\section{Projection-based Embedding Analytical Nuclear Gradients} 

\subsection{Projection-based Embedding Energy Theory} \label{sec:energy}
Projection-based WF-in-DFT embedding relies on the partitioning the LMOs of a system into two subsystems. Subsystem A contains the LMOs that are treated using the WF method and subsystem B contains the remaining LMOs that are treated using KS.
This WF-in-DFT procedure is accomplished by first performing a KS calculation on the full system to obtain a set of KS MOs.
The occupied KS MOs are then localized and partitioned into subsystems A and B.
Finally, subsystem A is treated using the WF method in the presence of the embedding potential created by the frozen LMOs of subsystem B.
Note that the cost of the KS calculation on the full system is typically negligible in comparison to the subsystem WF calculation. 
This results in our working equation for projection-based WF-in-DFT embedding, \cite{Manby2012}
\begin{equation} \label{eq:WFinDFTemb}
\begin{split}
E_{\text{WF-in-DFT}} & \sqbrak*{ \tilde{\Psi}^{\text{A}}; \gA,\gB } = E_{\text{WF}} \sqbrak*{\tilde{\Psi}^{\text{A}}} \\
    &+ \text{tr} \sqbrak*{ \paran{ \dA - \gA} \vembAB }  \\
	&+ E_{\text{DFT}} \sqbrak*{\gA+\gB} - E_{\text{DFT}} \sqbrak*{\gA} \\
	&+ \mu \text{tr} \sqbrak*{ \dA \mathbf{P}^{\text{B}} } \text{,} \\
\end{split}
\end{equation}
where $\tilde{\Psi}^{\text{A}}$ and $E_{\text{WF}}  \sqbrak{\tilde{\Psi}^{\text{A}}}$ are the WF and energy of subsystem A, $\dA$ is the subsystem A one-particle reduced density matrix that corresponds to $\tilde{\Psi}^{\text{A}}$, $E_{\text{DFT}}$ is the KS energy, and $\gA$ and $\gB$ are respectively the KS subsystem A and B one-particle densities that equal the full system KS density, $\boldsymbol{\gamma}$, when summed together. 
Throughout, we shall use a tilde to indicate  quantities  that have been calculated using the WF method. 
The embedding potential, $\vemb$, is defined as
\begin{equation} \label{eq:vemb}
\vembAB = \mathbf{g}\sqbrak*{\gA+\gB} - \mathbf{g}\sqbrak*{\gA} \text{,}
\end{equation} 
where $\mathbf{g}$ includes all KS two-electron terms, 
\begin{equation} \label{eq:g}
\paran*{ \mathbf{g} \sqbrak*{\boldsymbol{\gamma}} }_{\kappa \nu} = \sum_{\lambda \sigma} \gamma_{\lambda \sigma} \Big( (\kappa \nu | \lambda \sigma) - \frac{1}{2} x_f (\kappa \lambda | \nu \sigma) \Big) +  \paran*{ \mathbf{v}_{\text{xc}} [\boldsymbol{\gamma}] }_{\kappa \nu} \text{,}\\ 
\end{equation}
and where $\kappa$, $\nu$, $\lambda$ and $\sigma$ label atomic orbital basis functions, $(\kappa \nu| \lambda \sigma)$ are the two-electron repulsion integrals, $x_f$ is the fraction of exact exchange and $\mathbf{v}_{\text{xc}}$ is the exchange-correlation (XC) potential matrix.
The level-shift operator, $\mu \mathbf{P}^{\text{B}}$, is given by
\begin{equation}
\begin{split}
\mu \mathbf{P}^{\text{B}} = \mu \mathbf{S} \gB \mathbf{S} \text{,} \\
\end{split}
\end{equation}
where $\mathbf{S}$ is the overlap matrix. 
In the limit of $\mu \to \infty$, the LMOs that make up subsystems A and B are enforced to be exactly orthogonal, eliminating the non-additive kinetic energy present in other embedding frameworks \cite{Wesolowski1993,Gotz2009}. %
In practice, finite values of $\mu$ in the range of $10^4$~hartree to $10^7$~hartree are found to provide accurate results regardless of chemical system.\cite{Manby2012} 
If greater accuracy is needed, a perturbative correction outlined in Ref.~\citen{Manby2012} can be added to the WF-in-DFT energy expression to account for the finiteness of $\mu$, but in practice, this correction is found to contribute negligibly to the total energy and is thus neglected here.

Projection-based embedding can also be used for DFT-in-DFT embedding via a simplified version of Eq.~\ref{eq:WFinDFTemb}.
The working equation for projection-based DFT-in-DFT embedding is\cite{Manby2012}
\begin{equation} \label{eq:correctedDFTemb}
\begin{split}
E_{\text{DFT-in-DFT}} & \sqbrak*{ \newgA; \gA, \gB} = E_{\text{DFT}} \sqbrak*{ \newgA } \\
    & + \text{tr} \sqbrak*{ \paran*{ \newgA - \gA } \vembAB } \\
	& + E_{\text{DFT}} \sqbrak*{ \gA + \gB} - E_{\text{DFT}} \sqbrak*{\gA} \\
    & + \mu \text{tr} \sqbrak*{ \newgA \mathbf{P}^{\text{B}} } \text{.}\\
\end{split}
\end{equation}
The only differences between WF-in-DFT and DFT-in-DFT embedding is that the first term on the RHS of Eq.~\ref{eq:WFinDFTemb} is replaced with the KS energy on subsystem A, $E_{\text{DFT}} \sqbrak*{ \newgA }$, and in the second and last terms $\dA$ is reduced to the subsystem A KS density matrix, $\newgA$.
\subsection{Projection-based WF-in-DFT Embedding Gradient Theory} \label{sec:WFinDFT_gradient}
Since projection-based embedding is a non-variational theory, its analytical gradient is conveniently derived using a Lagrangian approach.
We  first construct a Lagrangian based on the projection-based WF-in-DFT energy.  We then minimize the Lagrangian with respect to the  variational parameters in the embedding energy, which include the subsystem A WF and the LMO coefficients.
Then we  show how to solve for each of the Lagrange multipliers and provide the working equation for the gradient of the total energy.

For consistency in notation, the MO coefficient matrix $\mathbf{C}$ refers to the entire set of KS MOs (occupied and virtual). 
The submatrix of $\mathbf{C}$ that refers to the (occupied) LMOs is denoted as $\mathbf{L}$ with column indices $i, j, k, l$.
The submatrix of $\mathbf{C}$ that refers to the canonical virtual space is denoted as $\mathbf{C}_{\text{v}}$ with column indices $a, b, c, d$. 
The indices $m, n, p, q$ are used to index generic molecular orbitals.

\subsection{Total Energy Lagrangian}
We now derive the total energy Lagrangian for projection-based WF-in-DFT embedding.
Where appropriate we will provide WF method specific examples (e.g. MP2) of general terms outlined in the equations.
The WF-in-DFT Lagrangian is
\begin{equation} \label{eq:MP2inDFT_Lag}
\begin{split}
&\mathcal{L} \left[\mathbf{C}, \tilde{\Psi}^{\text{A}}, \mathbf{\Lambda}, \mathbf{x}, \mathbf{z}^{\text{loc}}, \mathbf{z} \right] = \\
	& \quad E_{\text{WF-in-DFT}} \sqbrak*{ \tilde{\Psi}^{\text{A}}; \gA,\gB } + \sum_{s} \Lambda^{\text{WF,A}}_{s} c_{s} \\
	& + \sum_{pq} x_{pq} \paran*{ \mathbf{C}^{\dagger} \mathbf{S} \mathbf{C} - \mathbf{1} }_{pq} + \sum_{i>j} z_{i j}^{\text{loc}} r_{i j} \\
    & + \sum_{ai} z_{ai} \paran*{\mathbf{F} \sqbrak*{\gA + \gB} }_{ai} \text{ .}
\end{split}
\end{equation}
The first term on the right hand side (RHS) of Eq.~\ref{eq:MP2inDFT_Lag} is the projection-based WF-in-DFT embedding energy described by Eq.~\ref{eq:WFinDFTemb}.
The second term on the RHS of Eq.~\ref{eq:MP2inDFT_Lag} contains any constraints, $c_s$, and the corresponding Lagrange multipliers, $\Lambda^{\text{WF,A}}_{s}$, that arise from ensuring that the Lagrangian is variational with respect to parameters in the WF method.
The third term on the RHS constrains the KS MOs, $\mathbf{C}$, to be orthonormal, which accounts for the basis set being atom centered; this term is commonly referred to as the Pulay force  \cite{Pulay1969} %
and arises from the atomic orbital basis set being atom centered.
The localization conditions, $r_{i j} = 0$, take into account how the KS MOs are localized before being selected for subsystems A and B.
This is important because the LMOs will have a different dependence on nuclear perturbation than canonical MOs.
In this work, we use Pipek-Mezey localization \cite{Pipek1989} to obtain LMOs. 
Generalization to other localization methods (e.g. Boys \cite{Foster1960} and intrinsic bond orbitals\cite{Knizia2013}) is straightforward. 
The localization conditions for Pipek-Mezey are
\begin{equation} \label{eq:PipekMezey}
r_{i j} = \sum_{C} \paran*{ S_{ii}^{C} - S_{j j}^{C} } S^{C}_{i j} = 0 \quad \text{ for all } i>j \text{,}
\end{equation}
where $C$ corresponds to an atom in the molecule. 
The matrices $S^{C}$ are defined as
\begin{equation}
S^{C}_{k l} = \sum_{\alpha \in C} \sum_{\beta} \paran*{ L_{\alpha k} S_{\alpha \beta} L_{\beta l} + L_{\alpha l} S_{\alpha \beta} L_{\beta k} } \text{,}
\end{equation}
where the summation over $\alpha$ is restricted to basis functions at atom $C$.
The Brillouin conditions, $\paran*{ \mathbf{F} \sqbrak*{\gA + \gB}}_{ai}=0$, reflect how the KS MOs are optimized before being used to construct subsystems A and B. 
The Brillouin conditions are only needed because subsystem B is frozen at the KS level of theory. 
However, due to the non-additivity of the XC potential, the Lagrange multipliers, $\mathbf{z}$, span the full virtual-occupied space.

The type and number of constraints applied to the WF method depend on the chosen method.
For example, if the WF method is MP2 then the constraints are
\begin{equation} \label{eq:MP2_constraints}
\begin{split}
\sum_{s} \Lambda^{\text{MP2,A}}_{s} c_{s} = & \sum_{pq} \tilde{x}_{pq} \paran*{ \tilde{\mathbf{C}}^{\text{A} \dagger} \mathbf{S} \tilde{\mathbf{C}}^{\text{A}} - \mathbf{1} }_{pq} \\
    & + \sum_{ai} \tilde{z}_{ai} \paran*{ \mathbf{F}^{\text{A}} }_{ai} \big|_{i \in \text{A}} \text{,}
\end{split}
\end{equation}
where the first term on the RHS of Eq.~\ref{eq:MP2_constraints} constrains the Hartree-Fock MOs, $\tilde{\mathbf{C}}^{\text{A}}$, to be orthonormal, the condition $i \in \text{A}$ restricts the sum to occupied MOs in subsystem A, and the second term on the RHS are the Brillouin conditions using the embedded Fock matrix, $\mathbf{F}^{\text{A}}$. 
The embedded Fock matrix is defined as\cite{Manby2012}
\begin{equation}
    \mathbf{F}^{\text{A}} = \mathbf{h} + \mathbf{g} \sqbrak*{\newgA} + \vembAB + \mu \mathbf{P}^{\text{B}},
\end{equation}
where $\mathbf{h}$ is the standard one-electron Hamiltonian, $\mathbf{g}$ includes all of the usual HF two-electron terms and $\newgA$ is the subsystem A HF one-particle density. 
These constraints also arise in the derivation of the MP2 analytical nuclear gradient \cite{Schutz2004b}. %

For the projection-based WF-in-DFT energy to equal the Lagrangian, the Lagrangian must be minimized with respect to all of its parameters, including  $\tilde{\Psi}^{\text{A}}$, $\mathbf{C}$, and all of the Lagrange multipliers.

\subsection{Minimizing the Lagrangian with respect to the variational parameters of the WF method}
\label{subsection:WF_Lagrangian}
Upon minimizing the WF-in-DFT Lagrangian with respect to $\tilde{\Psi}^{\text{A}}$, only terms associated with the first two terms on the RHS of Eq.~\ref{eq:MP2inDFT_Lag} survive, all of which are familiar from the WF Lagrangian for the corresponding WF gradient theories.
\begin{equation} \label{eq:MP2inDFT_min_high}
\begin{split}
 \frac{\partial \mathcal{L}}{\partial \tilde{\Psi}^{\text{A}}} & = \frac{\partial E_{\text{WF}} \sqbrak*{\tilde{\Psi}^{\text{A}}} }{\partial \tilde{\Psi}^{\text{A}}} + \text{tr} \sqbrak*{ \frac{\partial \dA }{\partial \tilde{\Psi}^{\text{A}}} \vembAB } \\
 	&+ \mu \text{tr} \sqbrak*{ \frac{\partial \dA }{\partial \tilde{\Psi}^{\text{A}}} \mathbf{P}^{\text{B}} } +  \frac{\partial }{ \partial \tilde{\Psi}^{\text{A}} } \sum_{s} \Lambda^{\text{WF,A}}_{s} c_{s} = 0\\
\end{split}
\end{equation}
Since the embedding potential is independent of $\tilde{\Psi}^{\text{A}}$, the Z-vector coupled perturbed Hartree-Fock (Z-CPHF) equations of any post-HF method are only impacted through the eigenvalues of the subsystem A HF WF. 
Therefore, the solutions for the WF Lagrange multipliers (e.g. $\tilde{\mathbf{x}}$ and $\tilde{\mathbf{z}}$ for MP2 in Eq.~\ref{eq:MP2_constraints}) are obtained using the standard implementation of the WF gradient no matter what KS method is selected to describe subsystem B.
However, if an alternative embedding potential is used that depends on the subsystem A WF, such as the Huzinaga constraint (i.e. Ref.~\citen{Hegely2016}), then the formulation of the WF gradient is changed; the Z-CPHF equations for a general WF method would need to be modified to include the contributions from the derivative of the embedding potential with respect to the subsystem A WF, $\tilde{\Psi}^{\text{A}}$.

\subsection{Minimizing the Lagrangian with respect to the MO coefficients}
The remaining Lagrange multipliers, $\mathbf{z}^{\text{loc}}$, $\mathbf{x}$, and $\mathbf{z}$ in Eq.~\ref{eq:MP2inDFT_Lag}, are determined by minimizing the WF-in-DFT Lagrangian with respect to the variational parameters of the KS method, namely the MO coefficients, $\mathbf{C}$.
Differentiation of the Lagrangian with respect to these parameters yields
\begin{equation} \label{eq:MP2inDFT_min_low}
\begin{split}
\sum_{\mu} C_{\mu p} \frac{\partial \mathcal{L}}{\partial C_{\mu q}} & = E_{p q} + \paran*{ \mathbf{a} \sqbrak*{ \mathbf{z}^{\text{loc}} } }_{p q} \\
	& + \paran*{ \mathbf{D} \sqbrak*{ \mathbf{z} } }_{p q} + 2 x_{p q} = 0 \text{,}
\end{split}
\end{equation}
where
\begin{equation} \label{eq:MP2inDFT_E}
\begin{split}
E_{p q} = \sum_{\mu} C_{\mu p} \Bigg( & \frac{\partial E_{\text{WF-in-DFT}} \sqbrak*{ \tilde{\Psi}^{\text{A}}; \gA,\gB} }{\partial C_{\mu q}} \\
	&+ \frac{\partial }{\partial C_{\mu q}} \sum_{s} \Lambda^{\text{WF,A}}_{s} c_{s} \Bigg) \text{,}
\end{split}
\end{equation}
\begin{equation} \label{eq:MP2inDFT_azloc}
\begin{split}
\paran*{ \mathbf{a} \sqbrak*{ \mathbf{z}^{\text{loc}} } }_{p q} & = \sum_{\mu} C_{\mu p} \Bigg( \sum_{k>l} z_{k l}^{\text{loc}} \frac{\partial r_{k l}}{\partial C_{\mu q}} \Bigg) \\
	& = \sum_{k>l} \mathcal{B}_{pqkl} z_{k l}^{\text{loc}} \Big{|}_{q \in \text{occ}} \text{,}
\end{split}
\end{equation}
\begin{equation} \label{eq:MP2inDFT_Dz}
\begin{split}
& \paran*{ \mathbf{D} \sqbrak*{\mathbf{z}} }_{pq} = \sum_{\mu} C_{\mu p} \Bigg( \sum_{a k} z_{ak} \frac{\partial \paran*{ \mathbf{F} \sqbrak*{\gA + \gB} }_{ak}}{\partial C_{\mu q}} \Bigg) \\
	&= \sum_{ak} \mathcal{D}_{pqak} z_{ak} \\
    & = \paran*{ \mathbf{F} \sqbrak{\gA + \gB} \mathbf{z} }_{p q} \Big|_{q \in \text{occ}} \\
    & \quad + \paran*{ \mathbf{F} \sqbrak*{\gA + \gB} \mathbf{z}^{\dagger} }_{p q} \Big|_{q \in \text{vir} } + 2 \paran*{ \mathbf{V}[\bar{\mathbf{z}}] }_{p q} \Big{|}_{q \in \text{occ}} \text{,} \\
\end{split}
\end{equation}
and
\begin{equation} \label{eq:MP2inDFT_x}
2 x_{pq} = \sum_{\mu} C_{\mu p} \Bigg( \sum_{m n} x_{m n} \frac{\partial S_{m n}}{\partial C_{\mu q}} \Bigg) \text{.}
\end{equation}
The 4-dimensional tensors, $\mathcal{B}$ and $\mathcal{D}$, are expanded in Appendices \ref{appendix:PipekMezey} and \ref{appendix:MP2inDFT_orbital}, respectively, $\bar{\mathbf{z}}$ corresponds to $\mathbf{z} + \mathbf{z}^{\dagger}$, and $\mathbf{V}[\bar{\mathbf{z}}]$ includes all two-electron terms of the generalized Fock matrix and is shown explicitly in Appendix \ref{appendix:MP2inDFT_orbital}.
Since the embedded Fock matrix, $\mathbf{F}^{\text{A}}$, contains the embedding potential, $\vemb$, its derivative with respect to the MO coefficients, $\mathbf{C}$, is nonzero resulting in the WF relaxed density being needed to construct $\mathbf{E}$ in Eq.~\ref{eq:MP2inDFT_E}, which is explicitly shown in Appendix \ref{appendix:MP2inDFT_orbital}.
Therefore, the subsystem A WF gradient only affects the embedding contributions to the gradient through the WF relaxed density.

We now show that solving for the Lagrange multipliers leads to familiar coupled perturbed equations. 
Combining the stationary conditions described by Eq.~\ref{eq:MP2inDFT_min_low} with the auxiliary conditions $\boldsymbol{x} = \boldsymbol{x}^{\dagger}$ yields the linear Z-vector equations\cite{Schutz2004b}
\begin{equation} \label{eq:LinearZvec}
\paran*{ 1 - \mathcal{P}_{p q} } \paran*{\mathbf{E} + \mathbf{D} \sqbrak*{\mathbf{z}} + \mathbf{a} \sqbrak*{\mathbf{z}^{\text{loc}} } }_{p q} = 0 \text{,}
\end{equation}
where $ \mathcal{P}_{p q}$ permutes the indices $p$ and $q$, which is used to solve for $\mathbf{z}$ and $\mathbf{z}^{\text{loc}}$. 
The matrix $\boldsymbol{x}$ is then obtained as 
\begin{equation} \label{eq:DFTinDFT_overlap_lag}
x_{pq} = - \frac{1}{4} \paran*{ 1 + \mathcal{P}_{p q} } \paran*{ \mathbf{E} + \mathbf{D} \sqbrak*{ \mathbf{z}} + \mathbf{a} \sqbrak*{\mathbf{z}^{\text{loc}} } }_{p q} \text{.}
\end{equation}
The Lagrange multipliers $\mathbf{z}^{\text{loc}}$ pertain to the occupied-occupied MO space; considering only the  occupied-occupied part of Eq.~\ref{eq:LinearZvec} yields 
\begin{equation} \label{eq:LinearZvec_occ_occ}
\paran*{ 1 - \mathcal{P}_{i j} } \paran*{ \mathbf{E} + \mathbf{D} \sqbrak*{ \mathbf{z}} + \mathbf{a} \sqbrak*{ \mathbf{z}^{\text{loc}} } }_{i j} = 0 \text{.}
\end{equation}
Using the Brillouin conditions  
and the knowledge that $z_{a b} = z_{i j} = z_{i a} = 0$, Eq.~\ref{eq:LinearZvec_occ_occ} can be further simplified by showing that
\begin{equation} \label{eq:LinearZvec_brill}
\paran*{ 1 - \mathcal{P}_{i j} } \paran*{\mathbf{D} \sqbrak*{ \mathbf{z}} }_{i j} = 0 \text{.}
\end{equation}
The solutions, $\mathbf{z}^{\text{loc}}$, are thus independent of $\mathbf{z}$, such that Eq.~\ref{eq:LinearZvec_occ_occ} reduces to
\begin{equation}
E_{i j} - E_{j i} + \sum_{k > l} \paran*{ \mathcal{B}_{i j k l} - \mathcal{B}_{j i k l} } z^{\text{loc}}_{k l} = 0 \text{.}
\end{equation}
These are the Z-vector coupled perturbed localization (Z-CPL) equations, which are used to solve for $\mathbf{z}^{\text{loc}}$.
Subsequently, $\mathbf{a} \sqbrak*{ \mathbf{z}^{\text{loc}} }$ can be computed according to Eq.~\ref{eq:MP2inDFT_azloc}.

The Lagrange multipliers $\mathbf{z}$ pertain to the virtual-occupied MO space; considering only the  virtual-occupied part of Eq.~\ref{eq:LinearZvec} yields
\begin{equation}
\begin{gathered}
\paran*{ 1 - \mathcal{P}_{a i} } \paran*{\mathbf{E} + \mathbf{D} \sqbrak*{\mathbf{z}} + \mathbf{a} \sqbrak*{\mathbf{z}^{\text{loc}}} }_{a i} = 0 \text{,} \\
\end{gathered}
\end{equation}
which further simplifies to
\begin{equation} \label{eq:DFTinDFT_External_Occ}
\begin{split}
\big( \mathbf{E} + \mathbf{a} & \sqbrak*{ \mathbf{z}^{\text{loc}} } + \mathbf{F} \sqbrak*{\gA + \gB} \mathbf{z} - \mathbf{z} \mathbf{F} \sqbrak*{\gA + \gB} \\
	& + 2 \mathbf{V} \sqbrak*{ \bar{\mathbf{z}} }  \big)_{a i} = 0 \text{.} \\
\end{split}
\end{equation}
These are the Z-vector coupled perturbed Kohn-Sham (Z-CPKS) equations. 
Having solved the Z-CPL and Z-CPKS equations, the remaining Lagrangian multipliers associated with the orthogonality constraints, $\boldsymbol{x}$, can be obtained from Eq.~\ref{eq:DFTinDFT_overlap_lag}.

\subsection{Gradient of the Total Energy}
Once the Lagrangian is minimized with respect to all variational parameters, the gradient of the total energy takes the form
\begin{equation} \label{eq:MP2inDFT_grad}
\begin{split}
\frac{\text{d} E_{\text{WF-in-DFT}}}{\text{d} q} = \frac{\text{d} \mathcal{L}}{\text{d} q} & = \frac{\partial \mathcal{L}}{\partial q} + \frac{\partial \mathcal{L}}{\partial  \tilde{\Psi}^{\text{A}} } \frac{ \tilde{\Psi}^{\text{A}} }{\partial q} + \frac{\partial \mathcal{L}}{\partial \mathbf{C} } \frac{\partial \mathbf{C} }{\partial q} \\ 
	&= \frac{\partial \mathcal{L}}{\partial q}. %
\end{split}
\end{equation}
Since the Lagrangian is minimized with respect to the subsystem A WF and the KS LMO coefficients, calculation of the WF and KS LMO responses to nuclear perturbation, ${ \partial \tilde{\Psi}^{\text{A}} }/{\partial q} $ and ${ \partial \mathbf{C}}/{\partial q} $ respectively, are avoided.
This yields the following expression for the gradient,
\begin{equation} \label{eq:MP2inDFT_grad_MO}
\begin{split}
& E_{\text{WF-in-DFT}}^{(q)} = E_{\text{WF}}^{(q)} \sqbrak*{ \tilde{\Psi}^{\text{A}}} 
+  \sum_{\lambda \nu} \dA_{\lambda \nu} \paran{ \vemb^{(q)} }_{\lambda \nu} \\
	& + \mu \sum_{\lambda \nu} \dA_{\lambda \nu} \paran{ \mathbf{P}^{\text{B},(q)} }_{\lambda \nu}
	+ \sum_{s} \Lambda^{\text{WF,A}}_{s} c_{s}^{(q)} \\
    &- E_{\text{DFT}}^{(q)}[\gA] + E_{\text{DFT}}^{(q)}[\gA + \gB] 
    - \sum_{\lambda \nu} \gA_{\lambda \nu} \paran{ \vemb^{(q)} }_{\lambda \nu} \\
	& + \sum_{i > j} z_{i j}^{\text{loc}} r_{i j}^{(q)} + \sum_{ai} z_{ai} \paran*{\mathbf{F} \sqbrak*{\gA + \gB}}_{ai}^{(q)} + \sum_{mn} x_{mn} S_{mn}^{(q)} \text{,}
\end{split}
\end{equation}
where the superscript $(q)$ denotes the explicit derivative of the quantity with respect to a nuclear coordinate.
Eq.~\ref{eq:MP2inDFT_grad_MO} can be further  simplified by folding $\sum_{s} \Lambda^{\text{WF,A}}_{s} c_{s}^{(q)}$ into the first three terms on the RHS of Eq.~\ref{eq:MP2inDFT_grad_MO}, yielding
\begin{equation} \label{eq:MP2inDFT_grad_simp}
\begin{split}
&E_{\text{WF-in-DFT}}^{(q)} = E_{\text{WF}}^{q} \sqbrak*{ \tilde{\Psi}^{\text{A}}} 
+  \sum_{\lambda \nu} \paran{\dA_{\text{rel}}}_{\lambda \nu} \paran{ \vemb^{(q)} }_{\lambda \nu} \\
	& + \mu \sum_{\lambda \nu} \paran{\dA_{\text{rel}}}_{\lambda \nu} \paran{ \mathbf{P}^{\text{B},(q)} }_{\lambda \nu} \\
	& - E_{\text{DFT}}^{(q)}[\gA] + E_{\text{DFT}}^{(q)}[\gA + \gB] 
	- \sum_{\lambda \nu} \gA_{\lambda \nu} \paran{ \vemb^{(q)} }_{\lambda \nu} \\
	& + \sum_{i > j} z_{i j}^{\text{loc}} r_{i j}^{(q)} + \sum_{ai} z_{ai} \paran*{\mathbf{F}\sqbrak*{\gA  + \gB}}_{ai}^{(q)} + \sum_{mn} x_{mn} S_{mn}^{(q)} \text{.}
\end{split}
\end{equation}
Here, $E_{\text{WF}}^{q} \sqbrak{ \tilde{\Psi}^{\text{A}}}$ denotes the total derivative of the subsystem A WF energy with respect to nuclear coordinate, which can be directly calculated using existing WF gradient implementations, and $\dA_{\text{rel}}$ is the WF-relaxed density for subsystem A. 
For example, the MP2-relaxed density is
\begin{equation}
\dA_{\text{rel}} =
\dA
+ \tilde{\mathbf{C}}^{\text{A}} \tilde{\mathbf{z}} \tilde{\mathbf{C}}^{\text{A},\dagger}
= \newgA + \mathbf{d}^{(2)} + \tilde{\mathbf{C}}^{\text{A}} \tilde{\mathbf{z}} \tilde{\mathbf{C}}^{\text{A},\dagger} \text{,}
\end{equation}
which contains the subsystem A Hartree-Fock density, $\newgA$, the MP2 density matrix, $\mathbf{d}^{(2)}$, and the solutions of the subsystem A Brillouin conditions, $\tilde{\mathbf{C}}^{\text{A}} \tilde{\mathbf{z}} \tilde{\mathbf{C}}^{\text{A},\dagger}$. 
Eq.~\ref{eq:MP2inDFT_grad_simp} can be expressed in terms of the WF gradient on subsystem A and the derivative AO integrals, yielding our final expression for the projection-based WF-in-DFT analytical gradient,
\begin{equation} \label{eq:MP2inDFT_grad_AO}
\begin{split}
&E_{\text{WF-in-DFT}}^{(q)} = E_{\text{WF}}^{q} \sqbrak*{\tilde{\Psi}^{\text{A}}} \\
    &+ \text{tr} \sqbrak*{\mathbf{d}_{\text{a}} \mathbf{h}^{(q)} } + \text{tr} \sqbrak*{ \mathbf{X} \mathbf{S}^{(q)} } + \frac{1}{2} \sum_{\mu \nu \lambda \sigma} D_{\mu \nu \lambda \sigma} (\mu \nu | \lambda \sigma)^{(q)} \\
    &+ E_{\text{xc}}^{(q)} \sqbrak*{\boldsymbol{\gA} + \boldsymbol{\gB}} - E_{\text{xc}}^{(q)} \sqbrak*{\gA}\\
    &+ \text{tr} \sqbrak[\Big]{ \paran*{ \dA_{\text{rel}} - \gA } \paran*{ \mathbf{v}_{\text{xc}}^{(q)} \sqbrak*{ \boldsymbol{\gA} + \boldsymbol{\gB} } - \mathbf{v}_{\text{xc}}^{(q)} \sqbrak*{ \gA } } } \text{.}
\end{split}
\end{equation}
The effective one-particle density $\mathbf{d}_{\text{a}}$ and effective two-particle density $\mathbf{D}$ are defined 
\begin{equation}
\mathbf{d}_{\text{a}} = \gB + \mathbf{C} \mathbf{z} \mathbf{C}^{\dagger} \text{,}
\end{equation}
and
\begin{equation}
\begin{split}
D_{\mu \nu \lambda \sigma} &= \paran*{\gA + \gB}_{\mu \nu} \paran*{ \mathbf{d}_{\text{b}}}_{\lambda \sigma}  - \gamma^{\text{A}}_{\mu \nu} \paran*{ \mathbf{d}_{\text{c}} }_{\lambda \sigma} \\
	&- \frac{1}{2} x_{f} \paran*{ \paran*{\gA+\gB}_{\mu \lambda} \paran*{ \mathbf{d}_{\text{b}}}_{\nu \sigma} - \gamma^{\text{A}}_{\mu \lambda} \paran*{ \mathbf{d}_{\text{c}} }_{\nu \sigma} } \text{.} \\
\end{split}
\end{equation}
The effective one-particle densities $\mathbf{d}_{\text{b}}$ and $\mathbf{d}_{\text{c}}$ are defined 
\begin{equation}
\begin{split}
\mathbf{d}_{\text{b}} = \gA + \gB + 2 \mathbf{C} \mathbf{z} \mathbf{C}^{\dagger} + 2\dA_{\text{rel}} - 2\gA \text{,}
\end{split}
\end{equation}
and
\begin{equation}
\begin{split}
\mathbf{d}_{\text{c}} = - \gA + 2\dA_{\text{rel}} \text{.}
\end{split}
\end{equation}
The matrix $\mathbf{X}$ is defined 
\begin{equation}
\begin{split}
\mathbf{X} &= \mathbf{C} \mathbf{x} \mathbf{C}^{\dagger} + \sum_{i > j} \frac{ \partial r_{ij} }{ \partial S_{\mu \nu} } z_{ij}^{\text{loc}} \\
	 &= \mathbf{X}^{\text{loc}} - \frac{1}{2} \mathbf{L} \paran[\big]{ \mathbf{E} + 2 \mathbf{V} \sqbrak*{ \bar{\mathbf{z}} } } \mathbf{L}^{\dagger} \\
	&\quad - \frac{1}{2} \paran*{ \mathbf{C}_{\text{v}} \paran*{ \mathbf{z} \mathbf{F} } \mathbf{L}^{\dagger} + \paran*{ \mathbf{C}_{\text{v}} \paran*{ \mathbf{z} \mathbf{F} } \mathbf{L}^{\dagger} }^{\dagger} }\\
	&\quad + \mu \paran[\Big]{ \dA_{\text{rel}} \mathbf{S} \gB + \gB \mathbf{S} \dA_{\text{rel}} } \text{,}
\end{split}
\end{equation}
where 
\begin{equation} \label{eq:x_loc}
\paran*{ \mathbf{X}^{\text{loc}} }_{\mu \nu} = - \frac{1}{2} \paran*{ \mathbf{L} \mathbf{a} \sqbrak*{ \mathbf{z}^{\text{loc}} } \mathbf{L}^{\dagger} }_{\mu \nu} + \sum_{i > j} \frac{ \partial r_{ij} }{ \partial S_{\mu \nu} } z_{ij}^{\text{loc}} \text{.}
\end{equation}
The second term on the RHS of Eq.~\ref{eq:x_loc} is expanded in Appendix \ref{appendix:PipekMezey}.

The analytical nuclear gradient expression for projection-based DFT-in-DFT closely follows that for WF-in-DFT,  with regard to evaluation of both the Lagrange multipliers (Eq.~\ref{eq:MP2inDFT_min_low}) and the final gradient  (Eq.~\ref{eq:MP2inDFT_grad_AO}).
To obtain the corresponding DFT-in-DFT expressions,  $\dA_{\text{rel}}$ becomes the subsystem A KS density
\begin{equation}
\dA_{\text{rel}} = \newgA \text{,}
\end{equation}
which affects the evaluation of $\mathbf{E}$ in Eq.~\ref{eq:MP2inDFT_min_low} (expanded in Eq.~\ref{ap_eq:MP2inDFT_E}) and the evaluation of the final gradient expression, Eq.~\ref{eq:MP2inDFT_grad_AO}.
Additionally, the first term on the RHS of the final gradient expression, Eq.~\ref{eq:MP2inDFT_grad_AO}, is replaced with the subsystem A KS gradient, $E_{\text{DFT}}^{q} \sqbrak*{\newgA} $.

\section{Computational Details}
The implementation of projection-based WF-in-DFT embedding gradients is available in the 2019 general release of Molpro \cite{MOLPRO}. 
In all embedding calculations reported here, unless otherwise specified, the Pipek-Mezey localization method \cite{Pipek1989} is used with the core and occupied MOs localized together. 
The subsystem A region is chosen by including any LMOs with a net Mulliken population larger than 0.4 on the atoms associated with subsystem A, although more sophisticated partitioning algorithms have been introduced.\cite{Welborn2018}
A level-shift parameter of $\mu=10^6$~hartree is used for all embedding calculations.
The perturbative correction to using a finite value of $\mu$ in Eq.~\ref{eq:correctedDFTemb} is less than 20 microhartrees for the applications presented here and thus not included (accomplished by specifying the option $\texttt{HF\_COR} = 0$). 
Throughout this work, all embedding calculations are described using the nomenclature ``(WF method)-in-DFT/basis," where the WF method describes subsystem A and the KS method describes subsystem B.
For some embedding calculations a mixed-basis set is used and is denoted by ``(WF method)-in-DFT/large-basis:small-basis," where the large basis is used to describe subsystem A and the small basis is used to describe subsystem B.

All SCF calculations employ a tighter threshold than default for MO convergence by specifying the option $\texttt{ORBITAL} = 1 \times 10^{-7}$~a.u. in Molpro. 
All KS calculations used in projection-based embedding are done without density fitting, employing the local-density approximation (LDA) \cite{Hohenberg1964,Vosko1980}, Perdew-Burke-Ernzerhof (PBE) \cite{Perdew1996}, PBE0 \cite{Adamo1999a}, and LDAX functionals %
with the def2-TZVPP, def2-SVP, def2-ASVP, \cite{Weigend2005,Weigend2006} cc-pVDZ, \cite{Dunning1989} and 6-31G \cite{Hehre1972} basis sets. 
Note that the def2-ASVP basis set used in Molpro is constructed by adding one set of even tempered diffuse functions to the def2-SVP basis set. 
The LDAX functional is constructed by including $50\%$ exact exchange and reducing the weight of the DIRAC functional to $50\%$ in the LDA functional.
For the calculations in sections \ref{sec:comparison} and \ref{sec:co_neb}, the XC functional is evaluated on a fixed-pruned grid with index 7 (Ref.~\citenum{Neese2018}). 
For the optimized geometries shown in section \ref{sec:optgeom}, and the malondialdehyde calculations in section \ref{sec:malondialdehyde} the XC functional is evaluated on an adaptively generated quadrature grid that reproduces the energy of the Slater-Dirac functional to a specified threshold accuracy of $10^{-10} E_h$. 
All WF calculations are performed with the frozen-core approximation, without density fitting, employing the MP2 \cite{Møller1934}, coupled-cluster singles and doubles (CCSD) \cite{Scuseria1988,Purvis1982}, and coupled-cluster singles, doubles, and perturbative triples [CCSD(T)] \cite{Raghavachari1989} correlation treatments with the def2-TZVPP, def2-SVP, cc-pVDZ and 6-31G basis sets.
Even though the density fitting approximation is not used for the WF methods in this study, density fitted gradients are available for the aforementioned WF methods. \cite{Bozkaya2016,Bozkaya2017}
The default values for integral screening were used in Molpro.
For all Z-CPKS calculations an iterative subspace solver employing the Davidson algorithm \cite{Davidson1975,Kauczor2011} is used with a convergence threshold of $1 \times 10^{-6}$~a.u.
For all Z-CPHF calculations needed for the subsystem A WF gradient an iterative solver with a convergence threshold of $1 \times 10^{-7}$~a.u. is used.
Grid weight derivatives are included for all gradient calculations involving the XC functional and potential.

For all geometry optimizations the number of LMOs in subsystem A is kept unchanged throughout the optimization. 
A natural way of enforcing this in future work is to employ even-handed partitioning,\cite{Welborn2018} although this was not needed in the examples studied here; the default procedure based on net Mulliken population sufficed to keep subsystem A unchanged. 
All geometries are optimized using the translation-rotation-internal coordinate system devised by Wang and Song, \cite{Wang2016} %
which is available in the GeomeTRIC  package.\cite{GeomeTRIC} 
Convergence parameters for the geometry optimizations follow the default parameters used by Molpro, namely that the maximum gradient value becomes less than $3 \times 10^{-4}$~hartree/bohr and the energy change between adjacent steps becomes less than $1 \times 10^{-6}$~hartree or the maximum component of the step displacement becomes less than $3 \times 10^{-4}$~bohr.
The maximum gradient value is evaluated in the Cartesian basis. 
All geometries are provided in the supporting information. %

Nudged elastic band (NEB) \cite{Henkelman2000a} calculations are run using the  implementation of the method in the atomic simulation environment (ASE) package \cite{ase-paper}. %
All NEB calculations use Molpro forces which are provided through a Molpro calculator interface within the ASE package. 

The intramolecular proton transfer of malondialdehyde is modeled with an NEB consisting of 15 images connected by springs with spring constants of $0.1$ eV/\AA$^{2}$.
The CCSD/def2-aSVP, CCSD-in-LDA/def2-aSVP and LDA/def2-aSVP optimized NEBs used the image dependent pair potential (IDPP) method \cite{Smidstrup2014} as the initial guess for the band with reactant and product geometries previously optimized at the corresponding level of theory. 
All NEB calculations for malondialdehyde are converged with the Broyden-Fletcher-Goldfarb-Shanno (BFGS) update of the Hessian and by enforcing that the maximum gradient value is less than $0.01$ eV/\AA$^{2}$.

The intramolecular proton transfer of the organometallic cobalt complex is modeled with an NEB consisted of images connected by springs with spring constants of $9$~eV/\AA$^{2}$.
The PBE0/cc-pVDZ climbing image NEB \cite{Henkelman2000}, consisting of 26 images, used the IDPP method as the initial guess for the band with the reactant geometry previously optimized. 
The CCSD-in-PBE0/cc-pVDZ NEB consisting of 23 images, used its optimized reactant and the climbing image NEB converged at the PBE0/cc-pVDZ level of theory as its initial guess. 
Since the product is spatially far away from the reactant, an intermediate geometry between the transition state and the product is used as the endpoint of the NEB. 
This intermediate geometry is determined by initially converging a NEB with an extra image such that the maximum force dropped below $0.3$~eV/\AA$^{2}$. 
Then the second to last image is used as the new endpoint and a new NEB is converged. 
The PBE0/cc-pVDZ climbing image NEB is optimized using the FIRE \cite{Bitzek2006} algorithm using a convergence criteria of $0.05$ eV/\AA$^{2}$ for the maximum gradient value. 
The projection-based WF-in-DFT embedding NEB is optimized at the CCSD-in-PBE0/cc-pVDZ level of theory using the FIRE algorithm with a convergence criteria of $0.25$ eV/\AA$^{2}$ for the maximum gradient value. 

The CCSD-in-PBE0/cc-pVDZ calculations used for the NEB optimization are performed by specifying 51 occupied MOs to be in subsystem A using the \texttt{N\_ORBITALS} option and by using AO truncation with a threshold of $1~\times~10^{-3}$~a.u. 
An even-handed selection of AOs were used along the NEB by creating a union of the AOs that were selected for each image by the truncation procedure to ensure that the NEB traversed a smooth potential energy surface. 
The final, reported energies of the WF-in-DFT NEB are performed using the PNO-LCCSD\cite{Schwilk2015}/cc-pVDZ and PNO-LCCSD-in-PBE0/cc-pVDZ levels of theory. 
Both the PNO-LCCSD and PNO-LCCSD-in-PBE0 calculations are performed with density fitting using the cc-pVTZ/JKFIT \cite{Weigend2002a} (the def2-TZVPP/JKFIT \cite{Weigend2008} basis set was used for cobalt since the cc-pVTZ/JKFIT basis set was not available) and the cc-pVTZ/MP2FIT \cite{Weigend2002b} density fitting basis sets. 
Tighter domain approximations were employed for all PNO-LCCSD calculations by specifying the $\texttt{DOMOPT=TIGHT}$ option. 
Additionally, the Boughton-Pulay completeness criterion was used for the selection of the primary projected atomic orbitals domain by specifying the option $\texttt{THRBP=1}$ and the Pipek-Mezey localization method was used. 
For the PNO-LCCSD-in-PBE0 calculations, AO truncation is not used, the core and valence DFT molecular orbitals are localized separately using the Pipek-Mezey localization method, and the subsystem A orbitals are selected using the default procedure based on the Mulliken population threshold.  

All calculations using  AO truncation  \cite{Bennie2015} ensure that at least one AO is kept per atom (specified by option \texttt{AO\_PER\_ATOM}) to make evaluation of the integral derivative contributions from the one electron Hamiltonian simpler within Molpro.
This adds a negligible amount of AO functions than would have been selected using only the density threshold parameter \cite{Bennie2015} for the systems studied in this paper. 
In all embedding geometry optimizations that employ AO truncation, the number of truncated AOs is fixed using the \texttt{STOREAO} option to ensure smoothness of the potential energy function. 
Upon convergence, the truncated AO list is reevaluated using the same density threshold parameter; if the number of kept AOs remains a subset of the original list of truncated AOs then the optimization is converged.

\section{Results and Discussion} \label{sec:results}

\subsection{Comparison of Analytical and Numerical Gradients} \label{sec:comparison}

The implementation of the projection-based WF-in-DFT analytical gradient is tested by comparison with the gradient evaluated by numerical finite difference for a distorted geometry of ethanol. 
The finite difference gradients are evaluated using a 
four-point central difference formula with a base step size of $0.01$~bohr. 
The mean absolute error (MAE) between the analytical and finite difference gradients is reported for a range of embedding calculations in Table~\ref{tab:st_eth_MAE}. 
These results show that the analytical nuclear gradient for projection-based WF-in-DFT embedding is essentially numerically exact with respect to the gradients calculated by finite difference.  Comparison of the results obtained using HF over the full system versus using LDA over the full system illustrate that some of the finite difference error comes from the DFT exchange-correlation grid.  Comparison of the HF-in-HF results with full HF and of the LDA-in-LDA results with full LDA illustrate the modest effect of using a large-but-finite value for the level-shift operator in projection based embedding. 
These results confirm the correct implementation of projection-based WF-in-DFT analytical nuclear gradients.

\begin{table}
	\caption{Mean absolute error between the analytically and numerically determined embedding nuclear gradient for a distorted geometry of ethanol. The basis set 6-31G is used for all calculations. The distorted geometry of ethanol is provided in the supporting information.}
    \label{tab:st_eth_MAE}
	\begin{tabular*}{\columnwidth}{p{0.55\columnwidth} c}
	\hline
	Method 		& MAE (hartree/bohr) \\
	\hline
	HF      						& $5.00 \times 10^{-9}$ \\
	HF-in-HF						& $4.61 \times 10^{-8}$ \\
	LDA      						& $1.48 \times 10^{-8}$ \\
    LDA-in-LDA						& $7.23 \times 10^{-8}$ \\
    HF-in-LDA						& $5.24 \times 10^{-8}$ \\
    MP2-in-LDA						& $5.37 \times 10^{-8}$ \\
	CCSD-in-LDA						& $5.36 \times 10^{-8}$ \\
    CCSD(T)-in-LDA					& $5.26 \times 10^{-8}$ \\
    CCSD-in-LDA (AO)$^{\text{a}}$	& $3.48 \times 10^{-8}$ \\
    CCSD(T)-in-LDA (AO)$^{\text{a}}$& $3.40 \times 10^{-8}$ \\
    CCSD(T)-in-PBE0					& $5.26 \times 10^{-8}$ \\
    CCSD(T)-in-PBE0 (AO)$^{\text{a}}$& $1.12 \times 10^{-7}$ \\
	\hline
	\end{tabular*}
    ${\prescript{\text{a}}{}{}}$Calculations were performed with AO truncation with a density threshold of $1 \times 10^{-1} \text{a.u.}$
\end{table}

\subsection{Optimized Geometries} \label{sec:optgeom}

\subsubsection{Ethanol} \label{sec:ethanol}
\begin{figure}
    \includegraphics[width=0.5\columnwidth]{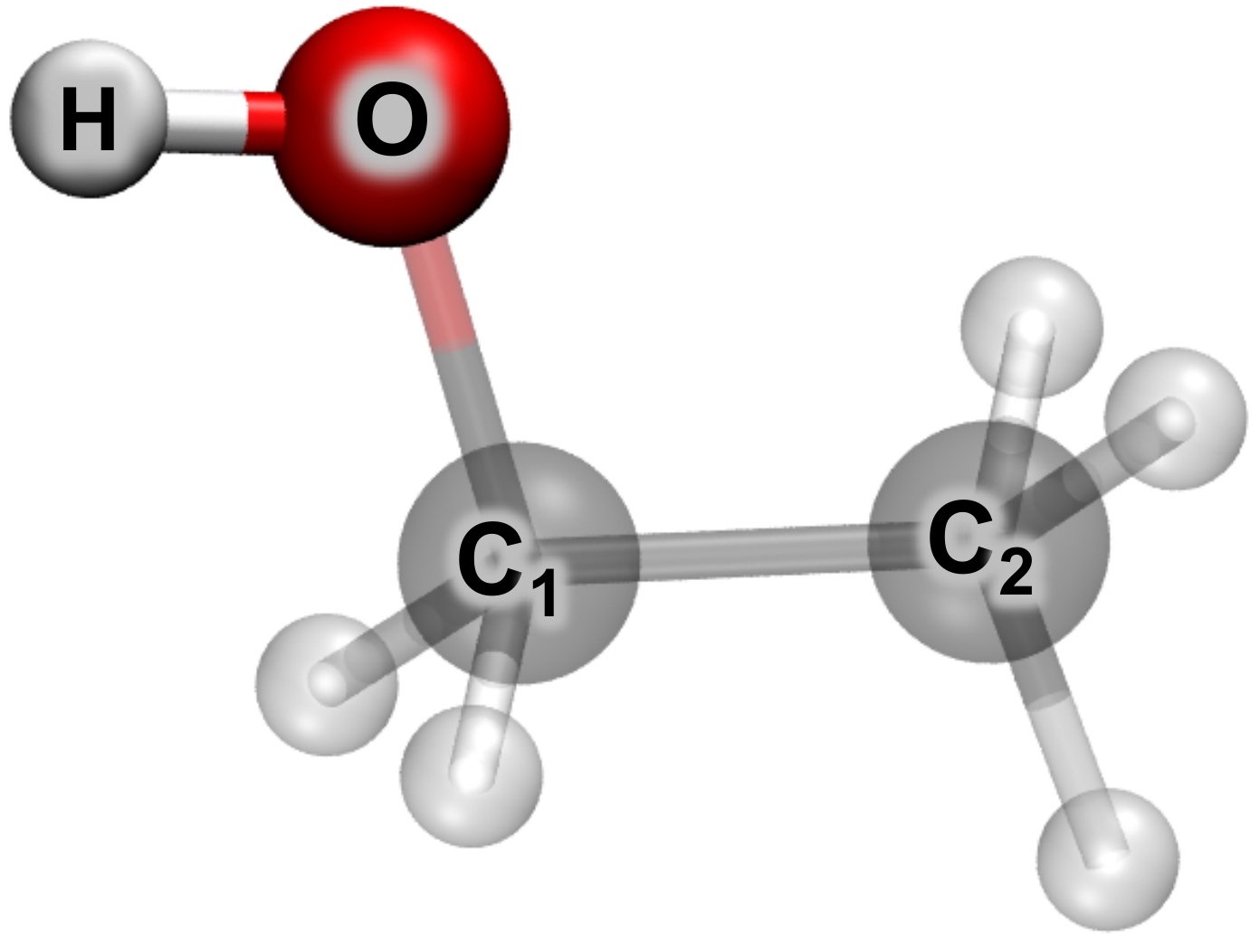}
    \caption{Optimized geometry for ethanol using projection-based CCSD-in-LDA/6-31G. 
    The solid atoms (O and H) are in subsystem A and the transparent atoms are in subsystem B.
    }
    \label{fig:opt_ethanol}
\end{figure}

As a proof of concept, CCSD-in-LDA/6-31G analytical nuclear gradients are employed to determine the ground state geometry of ethanol, which is shown in Fig.~\ref{fig:opt_ethanol}. 
For this simple case, the O-H moiety is treated by CCSD and the remainder of the molecule is treated by LDA.
Table~\ref{tab:opt_ethanol} shows that the O-H bond length within subsystem A reproduces the CCSD predicted bond length of $0.979$ \AA~and the remaining bonds within subsystem B reproduce the LDA predicted bond lengths.
This indicates that the potential energy surface produced by projection-based embedding varies smoothly from CCSD-like interactions for subsystem A and LDA-like interactions for subsystem B. 
Interestingly, the C-O bond located at the boundary between subsystems A and B closely reproduces the LDA bond length and is not an interpolation between the CCSD and LDA bond lengths.

\begin{table}
	\caption{Selected bond lengths and angles for ethanol (pictured in Fig.\ref{fig:opt_ethanol}) optimized at different levels of theory. Bond lengths are reported in units of Angstroms and angles are reported in units of degrees.}
    \label{tab:opt_ethanol}
    \begin{tabular}{lcccc}\hline
    Method 			& r(O-H)	& {$\angle$}$\text{C}_{1}$OH & r($\text{C}_{1}$-$\text{C}_{2}$) & r($\text{C}_{1}$-O)  \\ \hline
    LDA/6-31G 		& 0.988 	& 110.4 		             & 1.503 	                        & 1.439 \\
    CCSD-in-LDA/6-31G & 0.979 	& 110.7 		             & 1.506 	                        & 1.435 \\ 
    CCSD/6-31G 		& 0.979 	& 110.6 		             & 1.532 	                        & 1.475 \\ \hline
    \end{tabular}
\end{table}

\begin{figure}[H]
	\centering
    \includegraphics[width=1\columnwidth]{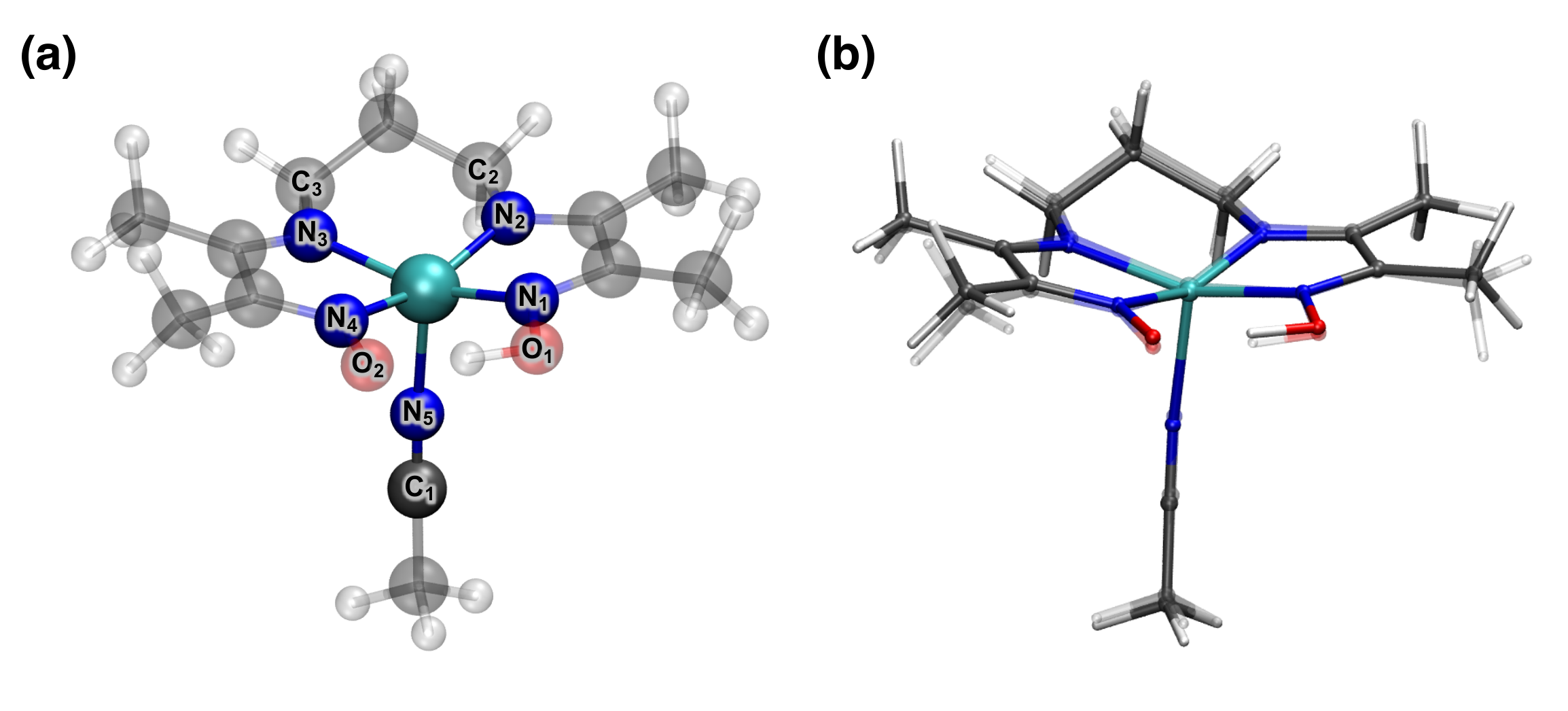}
    \caption{(a) The optimized geometry for the cobalt-based organometallic complex performed with projection-based CCSD-in-LDAX/def2-TZVPP:def2-SVP with AO truncation. The solid atoms (Co, N$_1$, N$_2$, N$_3$, N$_4$, N$_5$, and C$_1$) are included in subsystem A and the transparent atoms are included in subsystem B. 
    (b) The LDAX/def2-TZVPP:def2-SVP optimized geometry (transparent) and the projection-based CCSD-in-LDAX/def2-TZVPP:def2-SVP with AO truncation optimized geometry (solid). 
    }
    \label{fig:opt_co_I_CCSDinDFT}
\end{figure}

\subsubsection{Cobalt-based Organometallic Complex} \label{sec:co_neb}
As a demonstration of embedding gradients with AO truncation, the geometry of the cobalt-based organometallic complex, shown in Fig.~\ref{fig:opt_co_I_CCSDinDFT}, is optimized.
Fig.~\ref{fig:opt_co_I_CCSDinDFT}a shows the CCSD-in-LDAX/def2-TZVPP:def2-SVP optimized structure of the cobalt complex where the solid atoms are included in subsystem A and the transparent atoms are included in subsystem B.
In Fig.~\ref{fig:opt_co_I_CCSDinDFT}b the optimized structures evaluated at the CCSD-in-LDAX/def2-TZVPP:def2-SVP (solid) and the LDAX/def2-TZVPP:def2-SVP (transparent) levels of theory are overlaid.
While only modest differences are seen in the overall structure, Table~\ref{tab:opt_co_1} shows that the optimized bond lengths do change between the two levels of theory, both for the region within subsystem A and at the subsystem boundary. 
This indicates that the WF method is capable of relaxing the atoms in subsystem A even when they are strongly coordinated with subsystem B. It is also seen that the bond lengths across the boundary of subsystems A and B also differ from the LDAX geometry since the bonds in question experience the effects of both the WF and KS methods. 
Finally, if a bond length associated with atoms in subsystem B is considered, such as the O$_1$-H bond, it is found to closely match the LDAX predicted bond length. 

\begin{table}
	\caption{Selected bond lengths for the organometallic complex pictured in Fig.\ref{fig:opt_co_I_CCSDinDFT} optimized at different levels of theory and their absolute difference ($|\Delta|$). Bond lengths are reported in units of angstroms.}
    \label{tab:opt_co_1}
    \begin{tabular*}{0.9\columnwidth}{c|l@{\hskip 0.05\columnwidth}c@{\hskip 0.05\columnwidth}c@{\hskip 0.05\columnwidth}c}\hline
    &     		         & LDAX	        & CCSD-in-LDAX & $|\Delta|$   \\ \hline
    \parbox[t]{5mm}{\multirow{5}{*}{\rotatebox[origin=c]{90}{Sub A}}} &    r(Co-N$_1$)     & 1.836 	    & 1.846        & 0.010     \\
    &    r(Co-N$_2$)     & 1.893 	    & 1.883        & 0.010     \\
    &    r(Co-N$_3$)     & 1.932 	    & 1.951        & 0.019       \\
    &    r(Co-N$_4$)     & 1.900 	    & 1.926        & 0.026       \\
    &    r(Co-N$_5$)     & 1.978 	    & 2.026        & 0.048       \\ \hline
    \parbox[t]{5mm}{\multirow{5}{*}{\rotatebox[origin=c]{90}{Boundary}}} &    r(N$_5$-O$_1$)  & 1.317 	    & 1.355        & 0.038       \\ 
    &    r(N$_5$-O$_2$)  & 1.262 	    & 1.301        & 0.039       \\ 
    &    r(C$_1$-N$_5$)  & 1.131 	    & 1.150        & 0.019       \\ 
    &    r(N$_2$-C$_2$)  & 1.430 	    & 1.458        & 0.028       \\ 
    &    r(N$_3$-C$_3$)  & 1.428 	    & 1.458        & 0.030       \\ \hline
    \parbox[t]{5mm}{\multirow{3}{*}{\rotatebox[origin=c]{90}{Sub B}}} &&&& \\
    &    r(O$_1$-H)          & 1.025 	    & 1.030        & 0.005       \\ 
    &&&& \\ \hline
    \end{tabular*}
\end{table}

\begin{figure*}
    \includegraphics[width=0.6\textwidth]{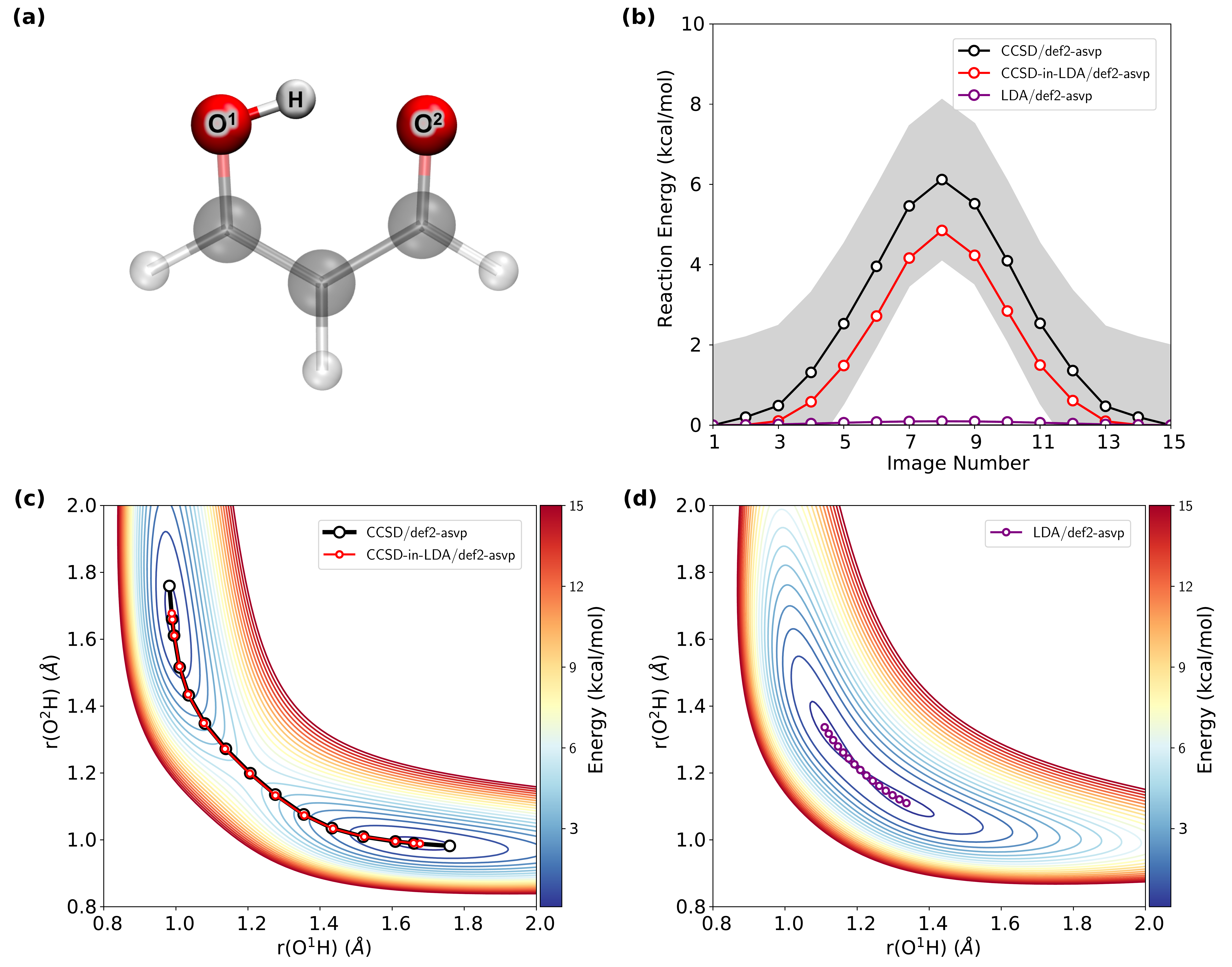}
    \caption{(a) The ground state geometry of malondialdehyde evaluated at the CCSD-in-LDA/def2-aSVP level of theory. The solid atoms are included in subsystem A and the transparent atoms are included in subsystem B. 
    (b) The reaction barrier heights for the minimum energy reaction pathways for LDA/def2-aSVP, CCSD-in-LDA/def2-aSVP and CCSD/def2-aSVP. (c), (d) Also shown are the minimum energy reaction pathways of the proton transfer in malondialdehyde as a function of the distance of the proton from the oxygen atoms, $\text{O}^{1}$ on the $x$-axis, $\text{O}^{2}$ on the $y$-axis for the CCSD-in-LDA/def2-aSVP and CCSD/def2-aSVP levels of theory, (c), and for the LDA/def2-aSVP level of theory, (d).}
    \label{fig:opt_mal_MERP}
\end{figure*}

\subsection{Malondialdehyde: Minimum Energy Reaction Pathway} \label{sec:malondialdehyde}
The minimum energy reaction pathway for the proton transfer in malondialdehyde is determined using the NEB method. 
Fig.~\ref{fig:opt_mal_MERP} shows that with minimal embedding (Fig.~\ref{fig:opt_mal_MERP}a) the CCSD-in-LDA/def2-aSVP reaction barrier, shown in Fig.~\ref{fig:opt_mal_MERP}b, is $4.85$ kcal/mol which is within $1.5$ kcal/mol of the CCSD/def2-aSVP reference reaction barrier of $6.12$ kcal/mol.
This is a vast improvement over the LDA/def2-aSVP result, which predicts an essentially barrierless reaction. 
In addition to correctly predicting the reaction barrier, Fig.~\ref{fig:opt_mal_MERP}c shows that the CCSD-in-LDA/def2-aSVP reaction pathway lies precisely on top of the CCSD/def2-aSVP pathway with only a small deviation in the basins.
In contrast, Fig.~\ref{fig:opt_mal_MERP}d shows that the LDA/def2-aSVP reaction pathway and potential energy surface reveal errors in the location of the reactant and product basins, with the hydrogen-bond length vastly underestimated. 
This is consistent with the tendency of LDA to over stabilize hydrogen bonds. 

\subsection{Cobalt-based Organometallic Complex: Minimum Energy Reaction Pathway}
\begin{figure*}
    \includegraphics[width=0.7\textwidth]{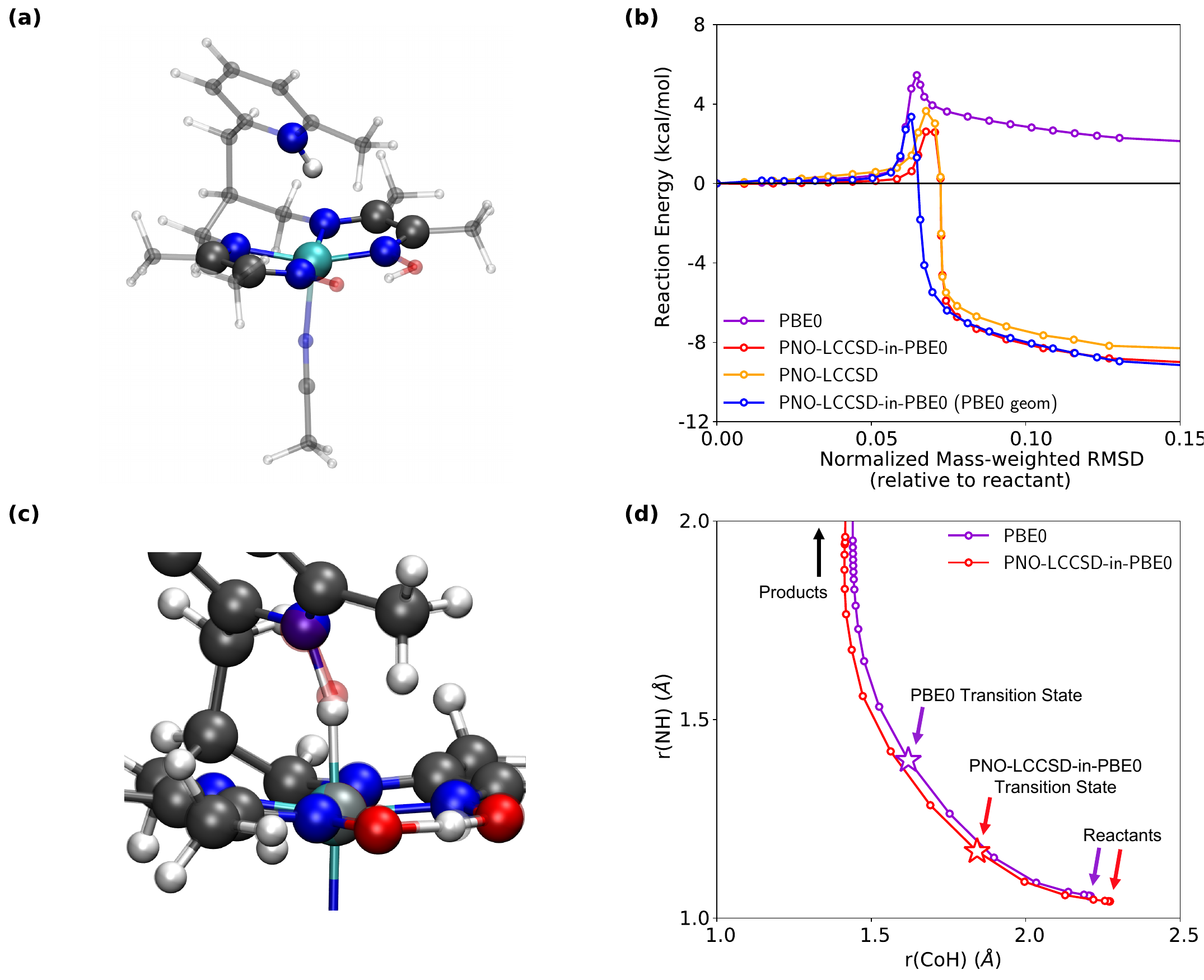}
    \caption{ All calculations used the cc-pVDZ basis set. 
    (a) The optimized geometry for a cobalt-based organometallic complex calculated at the CCSD-in-PBE0 level of theory with AO truncation. The solid atoms are included in subsystem A while the transparent atoms are in subsystem B.
    (b) The minimum energy reaction pathway for PBE0, the reaction pathway for PNO-LCCSD-in-PBE0 using the PBE0 geometries, the reaction pathway for PNO-LCCSD-in-PBE0 using CCSD-in-PBE0 geometries, and the reaction pathway for PNO-LCCSD using the CCSD-in-PBE0 geometries. The $x$-axis is a coordinate constructed by taking a normalized mass-weighted root-mean-square deviation (RMSD) of all images along the pathway with respect to the respective reactant and product. In comparing the purple and blue curves versus the orange and red curves, note that the transition state position in these normalized coordinates is affected by changes in the geometries of the reactant and product. 
    (c) A zoomed-in picture of the transition state geometries predicted by PBE0 (opaque atoms) and PNO-LCCSD-in-PBE0 (transparent atoms) levels of theory. The proton placement between the nitrogen and cobalt center at the PNO-LCCSD-in-PBE0 level of theory is highlighted in red.  
    (d) The minimum energy reaction pathways of the proton transfer for PNO-LCCSD-in-PBE0 and PBE0 as a function of the distance of the proton from the cobalt atom on the $x$-axis, and the nitrogen atom on the y-axis. The placement of the transition states are highlighted for each level of theory.
    }
    \label{fig:opt_co_MERP}
\end{figure*}

The minimum energy reaction pathway for the intramolecular proton transfer in a cobalt diimine-dioxime catalyst (Fig.~\ref{fig:opt_co_MERP}a) is now investigated. 
Previously, the reaction pathway for the transfer of the [-NH] to form a cobalt hydride had been investigated using geometries obtained using DFT. \cite{Huo2016}
Fig.~\ref{fig:opt_co_MERP}b shows the energy profile for this reaction determined by various levels of theory. 
We observe that the reaction pathway determined by the NEB optimized at the PBE0/cc-pVDZ level of theory (purple curve) predicts a barrier height of $5.45$~kcal/mol.
However, when single-point PNO-LCCSD-in-PBE0/cc-pVDZ embedding energy calculations are run on the PBE0 optimized geometries (blue curve), the barrier height is lowered to $3.35$~kcal/mol and the position of the transition state is shifted towards the reactant. 
The NEB optimized at the CCSD-in-PBE0/cc-pVDZ embedding level of theory (red curve) shows an even lower barrier height of $2.61$~kcal/mol and predicts a substantially different transition state geometry (Fig.~\ref{fig:opt_co_MERP}c) than the DFT result. 
The difference between the transition states predicted by the PBE0/cc-pVDZ and PNO-LCCSD-in-PBE0/cc-pVDZ levels of theory is clearly seen in Fig.~\ref{fig:opt_co_MERP}d, which shows the projection of the NEB onto the two dimensions of the Co-H and N-H bonds and with the position of the transition state geometry indicated with stars. 
This result clearly shows the large degree to which commonly employed DFT transition state geometries can differ from the CCSD-quality result that is obtained using projection-based embedding. 

\section{Conclusions}
We present the derivation and numerical demonstration of analytical nuclear gradients for projection-based embedding both with and without AO truncation. 
A key aspect of the gradient theory is that the WF contributions can be evaluated using existing WF gradient implementations without the need for modification or additional programming, thereby allowing projection-based WF-in-DFT embedding gradients to be easily generalized to any combination of WF and KS-DFT methods. 
It is demonstrated that projection-based embedding gradients produce accurate geometries for a variety of benchmark systems, including for bond-lengths that span the interface between subsystems. 
Furthermore, in applications to both malondialdehyde and a transition-metal catalyst, WF-in-DFT minimum energy pathways obtained via the NEB method reveal large errors in DFT-computed transition-state energies and geometries.  
Finally, we note that the Lagrangian framework presented here can be used to derive other analytical gradients of the projection-based WF-in-DFT energy with respect to quantities such as electric and magnetic fields.

\begin{acknowledgments}
We thank Matthew Welborn for helpful discussions. 
This material is based upon work supported by the U.S. Army Research Laboratory under Grant No. W911NF-12-2-0023 (S.J.R.L.). 
S.J.R.L. thanks the Caltech Resnick Sustainability Institute for a graduate fellowship. 
T.F.M. and F.R.M. acknowledge joint support from the DOE (Award No. DEFOA-0001912), and F.R.M. acknowledges support form the Engineering and Physical Sciences Research Council for funding (EP/M013111/1).

\end{acknowledgments}

\section{Supporting Information}
All geometries used in all tables and figures are available for download.

\appendix

\section{Pipek-Mezey Localization}
\label{appendix:PipekMezey}
Equation \ref{eq:MP2inDFT_azloc} from the main text 
\begin{equation}
\begin{split}
\paran*{ \mathbf{a} \sqbrak*{ \mathbf{z}^{\text{loc}} } }_{p q} & = \sum_{\mu} C_{\mu p} \Bigg( \sum_{k>l} z_{k l}^{\text{loc}} \frac{\partial r_{k l}}{\partial C_{\mu q}} \Bigg) \\
	& = \sum_{k>l} \mathcal{B}_{pqkl} z_{k l}^{\text{loc}} \Big{|}_{q \in \text{occ}}
\end{split}
\end{equation}
corresponds to the derivative of the localization conditions, Eq.~\ref{eq:PipekMezey}, with respect to the molecular orbital coefficients $\mathbf{C}$, where
\begin{equation}
\begin{split}
\mathcal{B}_{pqkl} \Big{|}_{q \in \text{occ}} = \sum_{C} & \Big[ \Big( 2 S^{C}_{p k} \delta_{k q} - 2 S^{C}_{p l} \delta_{l q} \Big) S^{C}_{k l} \\
    & + \Big( S^{C}_{k k} - S^{C}_{l l} \Big) \Big( S^{C}_{p l} \delta_{k q} + S^{C}_{p k} \delta_{l q} \Big) \Big] \text{.} \\
\end{split}
\end{equation}
Next, the overlap derivative contribution from the localization conditions from Eq.~\ref{eq:x_loc} is
\begin{equation}
\begin{split}
\sum_{i > j} \frac{ \partial r_{ij} }{ \partial S_{\mu \nu} } z_{ij}^{\text{loc}} &= \sum_{i > j} z_{ij}^{\text{loc}} \paran*{1 - \mathcal{P}_{ij}} \sum_{C} \Big[ 2 L_{\mu i} L_{\nu i} S^{C}_{i j} \\
    & + S^{C}_{ii} \paran*{ L_{\mu i} L_{\nu j} + L_{\mu j} L_{\nu i} }  \Big] \Big|_{\mu \in C} \text{,}
\end{split}
\end{equation}
where $\mathcal{P}_{ij}$ permutes the indices $i$ and $j$, and $\mu$ is restricted to atomic orbitals on atom $C$.

\section{Orbital Derivatives of Projection-based WF-in-DFT Embedding Energy}
\label{appendix:MP2inDFT_orbital}
This appendix provides additional details for the terms in Eqs.~\ref{eq:MP2inDFT_E} and \ref{eq:MP2inDFT_Dz} of the main text.
The derivative of the projection-based WF-in-DFT embedding energy and the WF constraints with respect to the MO coefficients shown in Eq.~\ref{eq:MP2inDFT_E} is
\begin{widetext}
\begin{equation} \label{ap_eq:MP2inDFT_E}
\begin{split}
E_{p q} &= \sum_{\mu} C_{\mu p} \Bigg( \frac{\partial E_{\text{WF-in-DFT}} \sqbrak*{ \tilde{\Psi}^{\text{A}}; \gA,\gB} }{\partial C_{\mu q}} + \frac{\partial }{\partial C_{\mu q}} \sum_{s} \Lambda^{\text{WF,A}}_{s} c_{s} \Bigg) \\
	& = \sum_{\mu} C_{\mu p} \Bigg( \frac{\partial E_{\text{DFT}} \sqbrak*{\gA + \gB} }{\partial C_{\mu q}} - \frac{\partial E_{\text{DFT}} \sqbrak*{\gA} }{\partial C_{\mu q}} + \text{tr} \sqbrak*{ \frac{\partial \paran*{\dA_{\text{rel}} - \gA} }{\partial C_{\mu q}} \vemb} + \text{tr} \sqbrak*{ \paran*{ \dA_{\text{rel}} - \gA} \frac{\partial \vemb}{\partial C_{\mu q}} } \\
    & \quad + \mu \text{tr} \sqbrak*{ \frac{\partial \dA_{\text{rel}} }{\partial C_{\mu q} } \mathbf{P}^{\text{B}} } + \mu \text{tr} \sqbrak*{ \dA_{\text{rel}} \frac{\partial \mathbf{P}^{\text{B}} }{\partial C_{\mu q} }} \Bigg) \text{,}
\end{split}
\end{equation}
where the partial derivative of the WF constraints causes the appearance of the WF relaxed density, $\dA_{\text{rel}}$, in the last four terms on the RHS of Eq.~\ref{ap_eq:MP2inDFT_E}.
Equation~\ref{ap_eq:MP2inDFT_E} simplifies to
\begin{equation} \label{ap_eq:MP2inDFT_E_2}
\begin{split}
E_{pq} & = 4 \paran*{ \mathbf{F} \sqbrak*{\boldsymbol{\gA} + \boldsymbol{\gB}} }_{pq} \Big{|}_{q \in \text{occ}} - 4 \paran*{ \mathbf{F} \sqbrak*{\gA} }_{pq} \Big{|}_{q \in \text{A}}  - 4 \paran*{\vemb}_{pq} \Big{|}_{q \in \text{A}} + 4 \paran*{ \mathbf{M} \sqbrak*{\dA_{\text{rel}} - \gA} }_{pq} \text{,} \\
\end{split}
\end{equation}
where $q \in \text{occ}$ indicates that the index $q$ is restricted to LMOs, $q \in \text{A}$ indicates that $q$ is restricted to LMOs in subsystem A, and $\mathbf{F}$ is the KS Fock matrix evaluated with the bracketed density.
The last term on the RHS of Eq.~\ref{ap_eq:MP2inDFT_E_2}
\begin{equation}
\paran*{ \mathbf{M} \sqbrak*{ \boldsymbol{\gamma} } }_{pq} = \frac{1}{4} \sum_{\mu} C_{\mu p} \paran*{ \text{tr} \sqbrak*{ \boldsymbol{\gamma} \frac{\partial \vembAB}{\partial C_{\mu q}} } + \mu \text{tr} \sqbrak*{ \boldsymbol{\gamma} \frac{\partial \mathbf{P}^{\text{B}} }{\partial C_{\mu q} }} } \text{,}
\end{equation}
simplifies to
\begin{equation}
\begin{split}
	\paran*{ \mathbf{M} \sqbrak*{ \boldsymbol{\gamma} } }_{p q} & = \sum_{\mu \nu} C_{\mu p} \sum_{\lambda \sigma} \gamma_{\lambda \sigma} \Big( (\mu \nu | \lambda \sigma) - \tfrac{1}{2} x_f (\mu \lambda | \nu \sigma) \Big) L_{\nu q}  \Big|_{q \in \text{B}} + \mu \paran*{ \mathbf{C}^{\dagger} \mathbf{P} \sqbrak*{ \boldsymbol{\gamma} } \mathbf{L} }_{p q} \Big|_{q \in \text{B}} \\
    & + \paran*{ \tilde{\mathbf{v}}_{\text{xc}} \sqbrak*{ \boldsymbol{\gA} + \boldsymbol{\gB}, \boldsymbol{\gamma} } }_{p q} \Big|_{q \in \text{occ}} + \paran*{ \tilde{\mathbf{v}}_{\text{xc}} \sqbrak*{ \boldsymbol{\gA}, \boldsymbol{\gamma} } }_{p q} \Big|_{q \in \text{A}} \\
\end{split}
\end{equation}
where 
\begin{equation}
\mathbf{P} \sqbrak*{ \boldsymbol{\gamma} } = \mathbf{S} \boldsymbol{\gamma} \mathbf{S} \text{.}
\end{equation}
In the current study, we employ both LDA and generalized gradient approximation (GGA) exchange-correlation functionals; for the special case of LDA,  the term $\tilde{\mathbf{v}}_{\text{xc}} \sqbrak*{ \gA + \gB, \boldsymbol{\gamma} }$ assumes the form
\begin{equation}
\begin{split}
\paran*{ \tilde{\mathbf{v}}_{\text{xc}} \sqbrak*{ \boldsymbol{\gA} + \boldsymbol{\gB}, \boldsymbol{\gamma} } }_{p q} &= \sum_{m n} \paran*{ pq | f_{\text{xc}} \sqbrak*{ \gA + \gB } | m n } \boldsymbol{\gamma}_{m n} \text{,} \\
\end{split}
\end{equation}
where $f_{\text{xc}}$ is the XC kernel which is defined as the second derivative of the XC functional with respect to density. 

The derivative of the Brillioun conditions in Eq.~\ref{eq:MP2inDFT_Dz} can be expanded as follows. 
\begin{equation}
\begin{split}
\paran*{ \mathbf{D} \sqbrak*{ \mathbf{z}} }_{pq} &= \sum_{ak} \mathcal{D}_{pqak} z_{ak} \\
	&= \sum_{a k} z_{a k} \Bigg[ \paran*{\mathbf{F}\sqbrak*{\gA + \gB}}_{p k} \delta_{a q} + \paran*{\mathbf{F}\sqbrak*{\gA + \gB}}_{a p} \delta_{k q} + 2 \sum_{l} \delta_{q l} \Big( 2 (a k | p l) - \tfrac{1}{2} x_f (a p | k l) - \tfrac{1}{2} x_f (a l | k p) \Big) \\
    & \qquad \qquad + \sum_{\mu \lambda \sigma} C_{\mu p} C_{\lambda a}  \frac{\partial \paran*{ \mathbf{v}_{\text{xc}} \sqbrak*{ \boldsymbol{\gA} + \boldsymbol{\gB} } }_{\lambda \sigma}}{\partial C_{\mu q}} L_{\sigma k} \Bigg] \\ 
    & = \paran*{ \mathbf{F} \sqbrak*{\gA + \gB} \mathbf{z} }_{p q} \Big|_{q \in \text{occ}} + \paran*{ \mathbf{F} \sqbrak*{\gA + \gB} \mathbf{z}^{\dagger} }_{p q} \Big|_{q \in \text{vir} } + 2 \paran*{ \mathbf{V} \sqbrak*{ \bar{\mathbf{z}} } }_{p q} \Big{|}_{q \in \text{occ}} \text{,}\\
\end{split}
\end{equation}
where $\bar{\mathbf{z}} = \mathbf{z} + \mathbf{z}^{\dagger}$ and $\mathbf{V}[\bar{\mathbf{z}}]$ is defined as
\begin{equation}
\begin{split}
& \paran*{ \mathbf{V} \sqbrak*{ \bar{\mathbf{z}} } }_{pq} \Big{|}_{q \in \text{occ}} = \sum_{m n} \bar{\mathbf{z}}_{m n} \Big( (m n | p q) - \tfrac{1}{2} x_f (m p | n q) \Big) \Big|_{q \in \text{occ}} + \paran*{ \tilde{\mathbf{v}}_{\text{xc}} \sqbrak*{ \boldsymbol{\gA} + \boldsymbol{\gB}, \bar{\mathbf{z}} } }_{p q} \Big|_{q \in \text{occ}} \text{.} \\
\end{split}
\end{equation}
\end{widetext}

\section{Atomic Orbital Truncation} \label{appendix:AO_energy}
Projection-based WF-in-DFT embedding reduces the cost of the WF calculation on subsystem A by reducing the number of LMOs that are correlated at the WF level, but thus far leaves the virtual space untouched.
However, the scaling of most WF methods is dominated by the number of virtual MOs (e.g. $\mathcal{O}(v^4)$ for CCSD).
One strategy has been to employ local correlation WF methods such as PNO-LMP2 \cite{Werner2015} and PNO-LCCSD \cite{Schwilk2017,Ma2018} to describe subsystem A 
since these methods are able to leverage the reduced number of LMOs to significantly lower the number of occupied-virtual orbital pairs that need to be included, resulting in a cheap and accurate WF calculation. 
However, a real advantage of projection-based embedding hinges on being able to use any WF method to describe subsystem A. 
Therefore, having a more general approach to reduce the cost of the WF calculation on the subsystem A is desirable.

The AO truncation scheme devised by Bennie et al. \cite{Bennie2015} provides a simple way to significantly reduce the cost of the WF calculation by reducing the size of the basis used to describe subsystem A.
The AOs that are discarded are selected through a single density threshold parameter: if the net Mulliken population, computed using the subsystem A density, of an AO is less than the specified threshold, it is removed from the basis set. 
This scheme has shown to greatly speedup up WF-in-DFT calculations at a small cost in accuracy in total and relative energies. \cite{Bennie2015}
Additionally, it has the nice feature that given a fixed subsystem A, the size of the truncated subsystem A basis scales asymptotically as the size of the environment grows.
This basis set modification does not cause any complications in the evaluation of the subsystem A WF gradient so existing implementations can be used without any modifications.
The energy expression for a projection-based WF-in-DFT calculation with AO truncation using the so-called type-in-type correction \cite{Bennie2015} is
\begin{equation} \label{eq:AO_WFinDFT}
\begin{split}
& E^{\text{trun}}_{\text{WF-in-DFT}} \sqbrak*{ \tilde{\Psi}^{\text{A,trun}}; \bar{\boldsymbol{\gamma}}^{\text{A,trun}} ; \gA, \gB } = E_{\text{WF}} \sqbrak*{\tilde{\Psi}^{\text{A,trun}}} \\
	&- E_{\text{DFT}}^{\text{trun}} \sqbrak*{ \bar{\boldsymbol{\gamma}}^{\text{A,trun}} } + E_{\text{DFT}} \sqbrak*{ \gA + \gB } \\
	& + \text{tr} \sqbrak*{ \paran*{\dAtrun - \bar{\boldsymbol{\gamma}}^{\text{A,trun}} } \vembABtrun } \\
	& + \text{tr} \sqbrak*{ \paran*{\dAtrun - \bar{\boldsymbol{\gamma}}^{\text{A,trun}} } \mathbf{P}^{\text{B,trun}} } \\
\end{split}
\end{equation}
where $\tilde{\Psi}^{\text{A,trun}}$ is the subsystem A WF in the truncated basis, $\bar{\boldsymbol{\gamma}}^{\text{A,trun}}$ is the KS subsystem A one-particle density in the truncated basis, $\gA$ and $\gB$ are the KS subsystem A and B one-particle densities in the full basis respectively, $\dAtrun$ is the subsystem A one-particle reduced density matrix that corresponds to $\tilde{\Psi}^{\text{A,trun}}$, $\vembABtrun$ is the embedding potential in the truncated basis which is evaluated by
\begin{equation}
	\vembABtrun = \mathbf{P}_{\text{t}}^{\dagger} \vembAB \mathbf{P}_{\text{t}} \text{,}
\end{equation}
and $\mathbf{P}^{\text{B,trun}}$ is the projection operator in the truncated basis which is evaluated by
\begin{equation}
	\mathbf{P}^{\text{B,trun}} = \mathbf{P}_{\text{t}}^{\dagger} \mathbf{P}^{\text{B}} \mathbf{P}_{\text{t}} \text{.}
\end{equation}
Here, $\mathbf{P}_{\text{t}}$ is the rectangular matrix that maps the full basis to the truncated basis which is created by starting with identity matrix and deleting columns corresponding to thrown away AO functions. 
We note that the even though the notation for Eq.~\ref{eq:AO_WFinDFT} is different from the one used in Ref.~\citenum{Bennie2015} the approach is identical. 

\section{Projection-based WF-in-DFT Gradient Theory with AO Truncation} \label{appendix:WFinDFT_gradient_AO}
\subsection{Total Energy Lagrangian}
We now derive the total energy Lagrangian for for projection-based WF-in-DFT embedding with AO truncation. 
The WF-in-DFT AO truncation Lagrangian is 
\begin{equation} \label{eq:DFTinDFT_Lag_AO}
\begin{split}
\mathcal{L} =& E{\mathstrut}^{\text{trun}}_{\text{WF-in-DFT}} \sqbrak*{ \tilde{\Psi}^{\text{A,trun}}; \bar{\boldsymbol{\gamma}}^{\text{A,trun}} ; \gA, \gB } + \sum_{s} \Lambda^{\text{WF,A}}_{s} c_{s} \\
	& - \sum_{ij \in \bar{\text{A}}} \bar{\epsilon}_{ij}^{\text{A}} \big[ \bar{\mathbf{C}}^{\text{A} \dagger} \mathbf{S} \bar{\mathbf{C}}^{\text{A}} - \mathbf{1} \big]_{ij} \\
    & + \sum_{p q} x_{p q} \big[ \mathbf{C}^{\dagger} \mathbf{S} \mathbf{C} - \mathbf{1} \big]_{pq}  + \sum_{i>j} z_{i j}^{\text{loc}} r_{i j} \\
    & + \sum_{ai} z_{ai} \paran*{ \mathbf{F} \sqbrak*{\gA + \gB}}_{ai} \text{,}
\end{split}
\end{equation}
where the bar superscript refers to subsystem A quantities optimized by the KS functional in the truncated basis.
The constraints that appear in Eq.~\ref{eq:DFTinDFT_Lag_AO} are all the same as those that appear in Eq.~\ref{eq:MP2inDFT_Lag} from the main text, except for the third term on the RHS of Eq.~\ref{eq:DFTinDFT_Lag_AO}.
This term constrains the MOs, $\bar{\mathbf{C}}^{\text{A}}$, to be orthogonal.

\subsubsection{Minimizing the Lagrangian with respect to the variational parameters of the WF method -- $\tilde{\Psi}^{\text{A,trun}}$.}
Minimizing the WF-in-DFT AO truncation Lagrangian with respect to $\tilde{\Psi}^{\text{A,trun}}$ simplifies to the minimization of the subsystem A WF energy and the WF constraints (as explained in section~\ref{subsection:WF_Lagrangian}), which corresponds to the conventional WF Lagrangian used to derive WF gradient theories, albeit in the truncated basis.

\subsubsection{Minimizing the Lagrangian with respect to the MO coefficients, ${\mathbf{\bar{C}}}^{\text{A}}$.}
The minimization of the Lagrangian with respect to the optimized KS MO coefficients in the truncated basis, ${\mathbf{\bar{C}}}^{\text{A}}$, results in the SCF equations using the embedded Fock matrix.
\begin{equation} 
\begin{split}
\sum_{\mu} \bar{C}^{\text{A}}_{\mu i} & \Bigg( \frac{\partial E^{\text{trun}}_{\text{DFT}}[ \bar{\boldsymbol{\gamma}}^{\text{A,trun}} ]}{\partial \bar{C}^{A}_{\mu j}} + \frac{\partial \text{tr} \sqbrak*{\bar{\boldsymbol{\gamma}}^{\text{A,trun}} \vembABtrun} }{\partial \bar{C}^{A}_{\mu j}} \\
	& + \mu \frac{\partial \text{tr} \sqbrak*{\bar{\boldsymbol{\gamma}}^{\text{A,trun}} \mathbf{P}^{\text{B,trun}} } }{\partial \bar{C}^{A}_{\mu j}} \Bigg) = \\
	&= \sum_{\mu} \bar{C}^{\text{A}}_{\mu i} \sum_{k l \in \bar{\text{A}}} \bar{\epsilon}^{\text{A}}_{k l} \frac{\partial \bar{S}^{A}_{k l}}{\partial \bar{C}^{\text{A}}_{\mu j}} \\
\end{split}
\end{equation}
\begin{equation} 
\begin{split}
\sum_{\mu \nu} \bar{C}^{A}_{\mu i} & \Big( \paran*{ \mathbf{F} \sqbrak*{\bar{\boldsymbol{\gamma}}^{\text{A,trun}}}}_{\mu \nu} +  \paran*{ \vembABtrun }_{\mu \nu} \\
	& + \mu \sqbrak*{ \mathbf{P}^{\text{B,trun}} }_{\mu \nu} \Big) \bar{C}^{A}_{\nu j} = \frac{1}{2} \bar{\epsilon}^{\text{A}}_{i j} \Big|_{ij \in \text{A}} \\
\end{split}
\end{equation}
\begin{equation} 
\begin{split}
\paran*{ \mathbf{F}^{\text{A}} \sqbrak*{\bar{\boldsymbol{\gamma}}^{\text{A,trun}}} }_{i j} \Big|_{ij \in \text{A}} = \frac{1}{2} \bar{\epsilon}^{\text{A}}_{i j} \Big|_{ij \in \text{A}}
\end{split}
\end{equation}
Therefore, the Lagrange multipliers $\frac{1}{2} \mathbf{\bar{\epsilon}^{\text{A}}}$ are simply the MO eigen energies of the KS optimized subsystem A MOs in the truncated basis.

\subsubsection{Minimizing the Lagrangian with respect to the MO coefficients, $\mathbf{C}$.}

The minimization of the Lagrangian with respect to the KS MO coefficients in the full basis, $\mathbf{C}$, is
\begin{equation} \label{eq:DFTinDFT_Zvec_AO}
\begin{split}
\sum_{\mu} C_{\mu p} \frac{\partial \mathcal{L}}{\partial C_{\mu q}} &= E_{p q} + \paran*{ \mathbf{a} \sqbrak*{ \mathbf{z}^{\text{loc}} } }_{p q}\\
	&+ \paran*{ \mathbf{D}\sqbrak*{\mathbf{z}} }_{p q} + 2 x_{p q} = 0 \text{.}
\end{split}
\end{equation}
where only the matrix $\mathbf{E}$ differs from the ones outlined in Eqns.~\ref{eq:MP2inDFT_E}-\ref{eq:MP2inDFT_x}.
\begin{equation}
\begin{split}
E_{pq} & = 4 \paran*{ \mathbf{F} \sqbrak*{\gA + \gB}}_{pq} \Big{|}_{q \in \text{occ}} \\
	& \quad + 4 \paran*{ \mathbf{M} \sqbrak*{ \mathbf{P}_{\text{t}} \paran*{ \dAtrun_{\text{rel}} - \bar{\boldsymbol{\gamma}}^{\text{A,trun}} } \mathbf{P}_{\text{t}}^{\dagger} } }_{pq} \\
\end{split}
\end{equation}
With the updated $\mathbf{E}$ matrix, the Lagrangian multipliers are solved in the same way as outlined for the WF-in-DFT Lagrangian multipliers.

\subsubsection{Gradient of the Total Energy}
Once the Lagrangian is minimized with respect to all variational parameters, the gradient of the energy with respect to nuclear coordinate, $q$, takes the form
\begin{equation} \label{eq:MP2inDFT_grad_AO_trun}
\begin{split}
&E_{\text{WF-in-DFT}}^{\text{trun},(q)} = E_{\text{WF}}^{\text{trun},q} \sqbrak*{\tilde{\Psi}^{\text{A,trun}}} - E_{\text{DFT}}^{\text{trun},q} \sqbrak*{ \bar{\boldsymbol{\gamma}}^{\text{A,trun}} } \\
    &+ \text{tr} \sqbrak*{\mathbf{d}_{\text{a}} \mathbf{h}^{(q)} } + \text{tr} \sqbrak*{ \mathbf{X} \mathbf{S}^{(q)} } + \frac{1}{2} \sum_{\mu \nu \lambda \sigma} D_{\mu \nu \lambda \sigma} (\mu \nu | \lambda \sigma)^{(q)} \\
    & + E_{\text{xc}}^{(q)} \sqbrak*{\boldsymbol{\gA} + \boldsymbol{\gB}} + \text{tr} \sqbrak[\Big]{ \mathbf{d}_{\text{c}} \paran*{ \mathbf{v}_{\text{xc}}^{(q)} \sqbrak*{ \boldsymbol{\gA} + \boldsymbol{\gB} } - \mathbf{v}_{\text{xc}}^{(q)} \sqbrak*{ \gA } } } \text{,}
\end{split}
\end{equation}
where the first two terms on the RHS of Eq.~\ref{eq:MP2inDFT_grad_AO_trun}, $E_{\text{WF}}^{\text{trun},q}$ and $E_{\text{DFT}}^{\text{trun},q}$, are the total derivative of the truncated subsystem A WF and KS energy respectively, minus the embedding contribution, $\vemb^{\text{trun,(q)}}$.
These two terms are calculated using existing gradient implementations, whereas the embedding contribution has been folded into the remaining terms.
The effective one-particle densities $\mathbf{d}_{\text{a}}$, $\mathbf{d}_{\text{b}}$ and $\mathbf{d}_{\text{c}}$ are 
\begin{equation}
\mathbf{d}_{\text{a}} = \boldsymbol{\gA} + \boldsymbol{\gB} + \mathbf{C} \mathbf{z} \mathbf{C}^{\dagger} \text{,}
\end{equation}
\begin{equation}
\begin{split}
\mathbf{d}_{\text{b}} = \boldsymbol{\gA} + \boldsymbol{\gB} + 2 \mathbf{C} \mathbf{z} \mathbf{C}^{\dagger} + 2 \mathbf{d}_{\text{c}} \text{,}
\end{split}
\end{equation}
and
\begin{equation}
\begin{split}
\mathbf{d}_{\text{c}} = \mathbf{P}_{\text{t}} \paran*{  \dAtrun_{\text{rel}} - \bar{\boldsymbol{\gamma}}^{\text{A,trun}} } \mathbf{P}_{\text{t}}^\dagger \text{.}
\end{split}
\end{equation}
The effective two-particle density $\mathbf{D}$ is 
\begin{equation}
\begin{split}
D_{\mu \nu \lambda \sigma} &= \paran*{ \boldsymbol{\gA} + \boldsymbol{\gB} }_{\mu \nu} \paran*{ \mathbf{d}_{\text{b}}}_{\lambda \sigma}  - 2 \gamma^{\text{A}}_{\mu \nu} \paran*{ \mathbf{d}_{\text{c}} }_{\lambda \sigma} \\
	&- \frac{1}{2} x_{f} \paran*{ \paran*{\boldsymbol{\gA} + \boldsymbol{\gB}}_{\mu \lambda} \paran*{ \mathbf{d}_{\text{b}}}_{\nu \sigma} - 2 \gamma^{\text{A}}_{\mu \lambda} \paran*{ \mathbf{d}_{\text{c}} }_{\nu \sigma} } \text{.} \\
\end{split}
\end{equation}
The matrix $\mathbf{X}$ is 
\begin{equation}
\begin{split}
\mathbf{X} &= \mathbf{X}^{\text{loc}} - \frac{1}{2} \mathbf{L} \paran[\big]{ \mathbf{E} + 2 \mathbf{V} \sqbrak*{ \bar{\mathbf{z}} } } \mathbf{L}^{\dagger} \\
	&- \frac{1}{2} \paran*{ \mathbf{C}_{\text{v}} \paran*{ \mathbf{z} \mathbf{F} } \mathbf{L}^{\dagger} + \paran*{ \mathbf{C}_{\text{v}} \paran*{ \mathbf{z} \mathbf{F} } \mathbf{L}^{\dagger} }^{\dagger} }\\ 
    &+ \mu \paran[\Big]{ \mathbf{d}_{\text{c}} \mathbf{S} \gB + \gB \mathbf{S} \mathbf{d}_{\text{c}} } \text{,}
\end{split}
\end{equation}
where 
\begin{equation} \label{eq:x_loc_trun}
\paran*{ \mathbf{X}^{\text{loc}} }_{\mu \nu} = - \frac{1}{2} \paran*{ \mathbf{L} \mathbf{a} \sqbrak*{ \mathbf{z}^{\text{loc}} } \mathbf{L}^{\dagger} }_{\mu \nu} + \sum_{i > j} \frac{ \partial r_{ij} }{ \partial S_{\mu \nu} } z_{ij}^{\text{loc}} \text{.}
\end{equation}

\bibliography{clean_Paper_Proj-Embed_Gradient}

\end{document}